\documentclass[5p,times]{elsarticle}
\usepackage{amssymb}
\usepackage{amsmath}
\usepackage[T1]{fontenc}
\usepackage[utf8]{inputenc} 
\usepackage[english]{babel}
\usepackage{microtype} 
\usepackage[shortcuts]{extdash}
\usepackage{cmap}
\usepackage{xcolor}
\usepackage{multirow}
\usepackage{booktabs}
\usepackage{graphicx}
\usepackage{listings}
\usepackage{tabularx}
\usepackage{booktabs}
\usepackage{longtable}
\usepackage{array}
\usepackage{cuted}
\usepackage{enumitem}
\usepackage{csquotes}
\usepackage{hyperref}
\usepackage{cleveref}
\usepackage{changes}
\usepackage{comment}
\usepackage{stfloats}
\newcolumntype{R}{>{\leavevmode\raggedleft\arraybackslash}X}
\makeatletter
\lst@AddToHook{TextStyle}{\let\lst@basicstyle\normalsize\ttfamily}
\makeatother

\crefname{lstlisting}{Listing}{Listings}
\Crefname{lstlisting}{Listing}{Listings}

\journal{Future Generation Computer Systems}

\newcommand{\var}[1]{\lstinline|#1|}

\begin{document}

\begin{frontmatter}

\title{Rethinking Sparse Formats for RISC-V: \\a Hierarchical Approach to High-Performance SpMV}

\author{Anna Pirova}
\author{Anastasia Vodeneeva}
\author{Konstantin Kovalev}
\author{Alexander Ustinov}
\author{Maksim Zagriadskov}
\author{Daniil Litvyakov}
\author{Arthur Kulik}
\author{Evgeny Kozinov}
\author{Valentin Volokitin}
\author{Iosif Meyerov\corref{cor1}}

\cortext[cor1]{Corresponding author.}
\ead{meerov@vmk.unn.ru}

\affiliation{organization={Department of HPC and System Programming, Lobachevsky State University of Nizhny Novgorod},
            addressline={23, Prospekt Gagarina}, 
            city={Nizhny Novgorod},
            postcode={603022}, 
            country={Russia}}

\begin{abstract}
The sparse matrix-vector multiplication (SpMV) algorithm is a fundamental computational kernel of linear algebra and serves as a building block for numerous applications, primarily iterative solvers for systems of linear equations used in scientific and engineering simulations. 
This paper compares vectorized implementations of the SpMV algorithm across eight established sparse matrix storage formats and proposes a novel modification of the CSR format, Hierarchical CSR (HCSR), which enhances SpMV performance on RISC-V processors. Our SpMV implementations utilize RVV 1.0 intrinsics and are publicly available as an open-source C++ library named RVVLASparse. 
Computational experiments conducted on SpacemiT K1 and K3 RISC-V boards demonstrate that the proposed HCSR format achieves the shortest execution time among all considered formats across a broad class of sparse matrices. HCSR outperforms the best CSR implementation on 8 cores, yielding an average speedup for double precision of 19\% on K1 (13\% on K3) and for single precision of 36\% on K1 (34\% on K3).
\end{abstract}




\end{frontmatter}

\section{Introduction}
\label{sectionIntro}

The emergence of the RISC-V open and patent-free instruction set architecture (ISA) has opened new research directions and opportunities for computer architects and high-performance software developers. Within just 15 years, the RISC-V ISA has evolved significantly from a research prototype to commercially available microprocessors and microcontrollers. Although the vast majority of currently produced devices belong to the IoT segment, early prototypes of server processors featuring dozens of modern out-of-order computational cores with RVV 1.0 vector extensions have emerged, making them suitable for many problems in the HPC domain \cite{33, 34}. The development of the RISC-V ecosystem requires optimization of the entire software stack, ranging from system-level and low-level software to end-user application packages. A potential solution is to rely entirely on an optimizing compiler, delegating code optimization responsibilities to it. This approach offers obvious advantages, including rapid adaptation speed, minimal effort expenditure, and the ability to simply recompile the code upon the release of newer, more advanced hardware. Nevertheless, a number of architecture-specific optimizations can still only be implemented manually, which entails substantial effort when adapting numerous applications.

The problem of mathematical software optimization for RISC-V is actively being investigated by the scientific and engineering communities. For instance, recent studies \cite{37, 39} examine an approach to optimizing codes in computational fluid dynamics and bioinformatics. In \cite{38}, the authors propose a method for RISC-V specific vectorization within the CatBoost machine learning package. In \cite{35}, performance issues are studied for astrophysics codes implemented using HPX and Kokkos, while \cite{43} proposes an optimization for the Fast Fourier Transform. Overall, the problem of tailoring mathematical software to the specifics of the new RISC-V ISA is of significant practical interest.

In this work, we investigate the dependence of sparse matrix-vector multiplication (SpMV) performance on the sparse matrix storage format for RISC-V processors. The SpMV algorithm is one of the key computational kernels in linear algebra. Specifically, it serves as the primary computationally intensive kernel for iterative solvers of sparse systems of linear equations (SLAE), which are applied to solve a wide range of engineering problems using supercomputers. Despite the apparent simplicity of this algorithm, efficient representation of sparse matrices that allows unlocking the potential of modern computing systems remains an unsolved challenge. Researchers proposed various approaches to increasing SpMV performance for x86 CPUs and GPUs through the use of specialized sparse matrix storage formats, leveraging SIMD instructions, and applying non-trivial parallelization schemes. The most common sparse linear algebra libraries for CPUs (e.g., Intel oneAPI MKL, AOCL-Sparse, SparseBLAS) utilize classic matrix storage formats that often fail to fully exploit the SIMD mechanisms supported by modern processors and are not adapted for RISC-V architectures. Meanwhile, most implementations of non-standard storage formats are distributed as standalone libraries and are not integrated into SLAE solvers.

Thus, although the problem is well-studied for traditional architectures, there remains room for further SpMV algorithm optimizations. These can be performed at different levels, ranging from leveraging the architectural peculiarities of the new RISC-V platform to attempting to overcome the fundamental memory wall issue, which prevents achieving acceptable performance in the SpMV operation (current results show a two-order-of-magnitude gap on the HPCG benchmark compared to Linpack test results).

In this paper, we present RVVLASparse \footnote{\href{URL}{https://github.com/UNN-ITMM-Software/RVVLASparse}}, a software library designed for RISC-V processors that enables the SpMV operation across nine different sparse matrix storage formats. Supported formats include CSR, Sell-C-$\sigma$, CSR2, CSR5, LAV, VHCC, VNEC, CVR, as well as a novel format, Hierarchical CSR (HCSR). The core idea behind the new format lies in improving memory subsystem efficiency by partitioning matrices into blocks of CSR and COO formats. 

The scientific contributions of this work are as follows:
\begin{enumerate}[itemsep=2pt, parsep=0pt, topsep=2pt, partopsep=0pt]
    \item We propose a new sparse matrix storage format, HCSR. Compared to scalar and typical vectorized SpMV implementations for the CSR format, the new format improves average SpMV performance on RISC-V CPUs by an average of 10\% for double precision and 30\% for single precision.
    \item We present a comparative performance analysis of the SpMV operation on two generations of RISC-V processors across different storage formats.
    \item We discuss the tunability of hyperparameters for various storage formats and identify the main characteristics of sparse matrices that influence format selection.
    \item We release the RVVLASparse library, which implements SpMV algorithms for nine sparse data structures for RISC-V CPUs, as publicly available open-source software.
\end{enumerate}

The remainder of this paper is structured as follows. 
Section \cref{sectionRW} provides an overview of sparse matrix storage formats. Section \cref{sectionAlgs} describes several matrix storage formats along with their corresponding SpMV algorithms and details the new HCSR format. Section \cref{sectionExp} presents experimental results for all implemented formats and analyzes SpMV performance. Section \cref{sectionML} describes the results of automatic storage format selection based on matrix sparsity pattern characteristics using gradient boosting of decision trees. Finally, Section \cref{secConcl} concludes the paper and outlines future work plans.

\section{Related Work}
\label{sectionRW}

Several papers on software development for RISC-V CPUs address low-level optimization of linear algebra algorithms. 
Specifically, we previously proposed efficient vectorized implementations for BLAS operations with band matrices based on the reference from OpenBLAS \cite{36}. In \cite{42}, dense linear algebra algorithms in the Eigen library were optimized for the RISC-V architecture. In \cite{45, 46, 47}, new vector instructions improved the performance of linear algebra operations.

SpMV is a widely studied memory-bound problem. Numerous papers propose improving SpMV performance on CPUs and GPUs by using non-standard sparse matrix storage formats. The classical sparse matrix storage formats such as Compressed Sparse Row (CSR), compressed column storage (CCS), and coordinate format (COO) were developed in the 1960s–1970s and remain the standard, most widely used data structures for general sparse matrices. Since 2000, many modifications of the classical storage formats have been developed, primarily targeting x86 processors with support for vector SIMD instructions and GPUs. A detailed survey of sparse matrix storage formats can be found, for example, in \cite{15}. Below, we discuss some of them.

Many data structures are built as modifications of the CSR format by grouping nonzero elements into dense blocks or lanes of equal size, which allows better utilization of vector operations and improves cache locality compared to standard CSR. These formats include CSR5 \cite{8}, CSR2 \cite{9}, CVR \cite{12}, and others. Such formats demonstrate performance gains over CSR on general sparse matrices.
Other modifications of CSR aim to improve SpMV performance on matrices with regular structure. For example, the CSX format \cite{18} extracts regular dense patterns (e.g., dense rectangular blocks, diagonal and vertical lines) within the matrix, while the VBFS format \cite{14} uses dense blocks of non-fixed size.
For matrices with small dense blocks, block analogues of classical formats are often used: block compressed sparse row (BCSR), block coordinate (BCOO), and generalized block coordinate (BCCOO) \cite{2}. Many papers have investigated hybrid versions of classical formats in various combinations, for example, COO~+~ELL (HYB format \cite{7}), COO~+~CSR, CSR~+~DIA, SELL-C-$\sigma$~+~CSR5 (HYB5 \cite{16}), and others. For GPUs, modifications of the ELLPACK format \cite{3} are widely used: ELL-R \cite{4}, SELL \cite{5}, SELL-C-$\sigma$ \cite{6}, and others. The latter format, SELL-C-$\sigma$, also demonstrates high performance on x86 processors \cite{17}. In \cite{41}, the performance of this format was studied on CPUs supporting long vector registers.

Note that for the x86 architecture, the most common software libraries (Intel oneAPI MKL, AOCL-Sparse, SparseBLAS, and others) support sparse matrix representations in CSR, CSC, COO, DIA, and BCSR formats. For graphics processors, CSR, BCSR, CSC, COO, and ELL modifications are typically supported. For most custom storage formats, the SpMV operation is implemented as a standalone code. Although often publicly available, integrating it into third-party software packages, such as iterative SLAE solvers, frequently requires significant additional engineering effort.

A number of papers discuss automatic selection of the optimal matrix storage format to yield the best performance using machine learning methods. For instance, the paper \cite{28} proposed the WISE machine learning framework, which uses decision trees to predict the SpMV operation speedup for various storage formats compared to CSR. In \cite{30}, CNNs trained on 2D matrix images were utilized. The authors of \cite{31} proposed an approach for determining the optimal threading configuration for SpMV computations in shared-memory architectures. In \cite{32}, a method is proposed for generating a machine-designed format and SpMV algorithm that combines several base formats. Our project adapts ideas from the WISE framework for the mechanism of automatic format selection based on the matrix structure.

We also note that a closely related work \cite{40} appeared during the preparation of this paper. In \cite{40}, vectorized algorithms for the SpMV operation were proposed for CSR, SELL-p, and JDS formats on RISC-V processors. The results were presented on three RISC-V platforms and demonstrated a noticeable performance improvement for SpMV due to vectorization. The best performance was achieved by the authors for the SELL-p format. A comparison with this implementation will be presented in Section \cref{sectionAnalysis}.

\section{SpMV Algorithm for Various Matrix Storage Formats}
\label{sectionAlgs}

\subsection{Sparse matrix-vector multiplication}

Let $A \in \mathbb{R}^{N \times M}$ be a sparse matrix with $NZ$ nonzeros, $b \in \mathbb{R}^{M}$ and $y \in \mathbb{R}^{N}$ be dense vectors, and $\alpha, \beta \in \mathbb{R}$ be scalar coefficients. Real numbers are represented as floating-point numbers. The objective is to compute the result of the following operation: 
\begin{equation}
\label{SpMV_operation}
y = \alpha {A b}+\beta y
\end{equation}

By definition, every element of the vector $y$ is calculated as follows:
\begin{equation}
\label{SpMV_dense}
y_i = \sum_{j=1}^{M} \alpha A_{ij} b_j + \beta y_j \quad \text{for} \quad i = 1, \dots, N
\end{equation}

\begin{figure*}[!thb]
\centering
\includegraphics[width=0.8\textwidth]{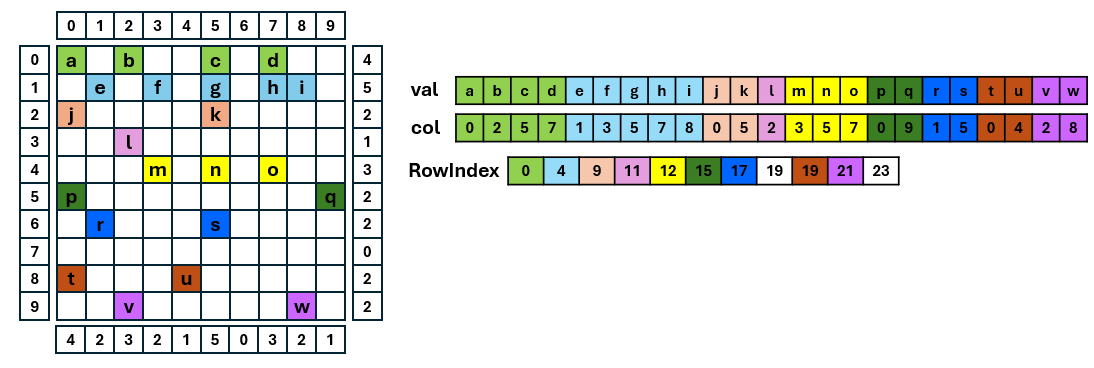}
\caption{Sparse matrix storage scheme using the CSR format.}
\label{csr}
\end{figure*}

For a sparse matrix $A$, the algorithm for evaluating \cref{SpMV_dense} depends on the matrix storage format. A straightforward implementation of SpMV using the traditional CSR, CCS, and COO formats has significant performance limitations.

In this paper, we present an implementation of the SpMV operation in both single and double precision using nine storage formats: CSR, Sell-C-$\sigma$, CSR2, CSR5, CVR, VHCC, VNEC, LAV, and a novel format introduced herein, named Hierarchical CSR (HCSR). The implementation is built upon RVV 1.0 intrinsics and OpenMP. For the sake of conciseness and clarity, we restrict our algorithmic exposition to the formats that achieved the highest performance in our numerical evaluation, along with a thorough description of the proposed HCSR format. Detailed discussions of the other formats are deferred to their original publications.

\subsection{Performance problems}
The SpMV algorithm is known to perform poorly on modern microprocessor architectures. Before presenting sparse matrix formats, we briefly outline the primary factors limiting SpMV performance. Many of these bottleneck issues are not intrinsic to the CSR or COO formats themselves; rather, they stem from the nature of sparse matrix representation and constitute inherent challenges in this domain. First, the memory footprint of sparse matrices and the access patterns of the SpMV algorithm result in irregular memory access to vector $\mathbf{b}$, causing frequent cache misses, poor cache utilization, and reduced data reuse. To address this, we propose the HCSR format, which can mitigate these adverse effects under certain conditions. Second, vectorizing the SpMV algorithm encounters fundamental performance degradation due to computationally expensive indexed loads. This effect is severe enough that vectorization can actually slow down execution for many matrices. Consequently, we evaluate the scalar and vectorized implementations of the CSR format separately. 
Third, achieving efficient parallelization is complicated by matrix-dependent optimal load balancing schedules. Thus, we empirically select scheduling strategies that deliver reasonable average performance across a broad matrix dataset.

\subsection{CSR Format}
The Compressed Sparse Row (CSR) format is a standard sparse matrix storage scheme supported by most sparse linear algebra libraries. In CSR, the nonzeros of the matrix are stored row by row. The representation consists of three arrays: \var{Val} stores the nonzero values in row-major order, \var{Col} holds their column indices, and \var{rowIndex} marks the beginning of each row within the \var{Col} array. The \var{Val} and \var{Col} arrays are of size \var{NZ}, and \var{rowIndex} is of size \var{N + 1} (see \cref{csr}). 

Listing \ref{alg:CSR} shows the pseudo-code for the vectorized implementation of CSR-based SpMV in double precision. The algorithm assumes that the vector register holds \var{gvl} double-precision elements.

\begin{figure*}[!thb]
\begin{lstlisting} [caption={SpMV algorithm for CSR format}, label={alg:CSR}]
void SP_MV_CSR_RVV(CSRmatrix mat, double* b, double* y, double alpha, double beta) {
	size_t vlmax = __riscv_vsetvlmax_e64m4();
	#pragma omp parallel for
	for (int i = 0; i < mat.m; ++i) {
		size_t vl = vlmax;
		vfloat64m4_t res = __riscv_vfmv_v_f_f64m4(0.0, vl);
		for (int j = mat.Rst[i]; j < mat.Rst[i + 1]; j += vlmax) {
			vl = __riscv_vsetvl_e64m4(mat.Rst[i + 1] - j); { /*@\label{line:crs1}@*/
			vfloat64m4_t val = __riscv_vle64_v_f64m4(mat.Val + j, vl);
			vuint32m2_t index = __riscv_vle32_v_u32m2(mat.Col + j, vl);
			vuint32m2_t index_shift = __riscv_vsll_vx_u32m2(index, 3, vl);
			vfloat64m4_t b_ = __riscv_vloxei32_v_f64m4(b, index_shift, vl); /*@\label{line:crs2}@*/
			res = __riscv_vfmacc_vv_f64m4_tu(res, val, b_, vl); /*@\label{line:crs3}@*/
		}
		vl = __riscv_vsetvl_e64m4(vlmax);
		vfloat64m1_t zero = __riscv_vfmv_v_f_f64m1(0.0, vl); /*@\label{line:crs4}@*/
		vfloat64m1_t sum = __riscv_vfredusum_vs_f64m4_f64m1(res, zero, vl);
		double tmp = __riscv_vfmv_f_s_f64m1_f64(sum); /*@\label{line:crs5}@*/
		y[i] = alpha * tmp + beta * y[i];
	}
}
\end{lstlisting}
\end{figure*}

In the pseudo-code, the dot product of row \var{A[i]} and vector \var{b} is computed via a contiguous vector load from \var{Val} and an indexed load from \var{b} (lines \ref{line:crs1}--\ref{line:crs2}). The products are accumulated in \var{res} (line \ref{line:crs3}) and subsequently reduced after the loop (lines \ref{line:crs4}--\ref{line:crs5}).

\subsection{Sell-C-$\sigma$ Format}
\begin{figure*}[!tb]
\centering
\includegraphics[width=0.8\textwidth]{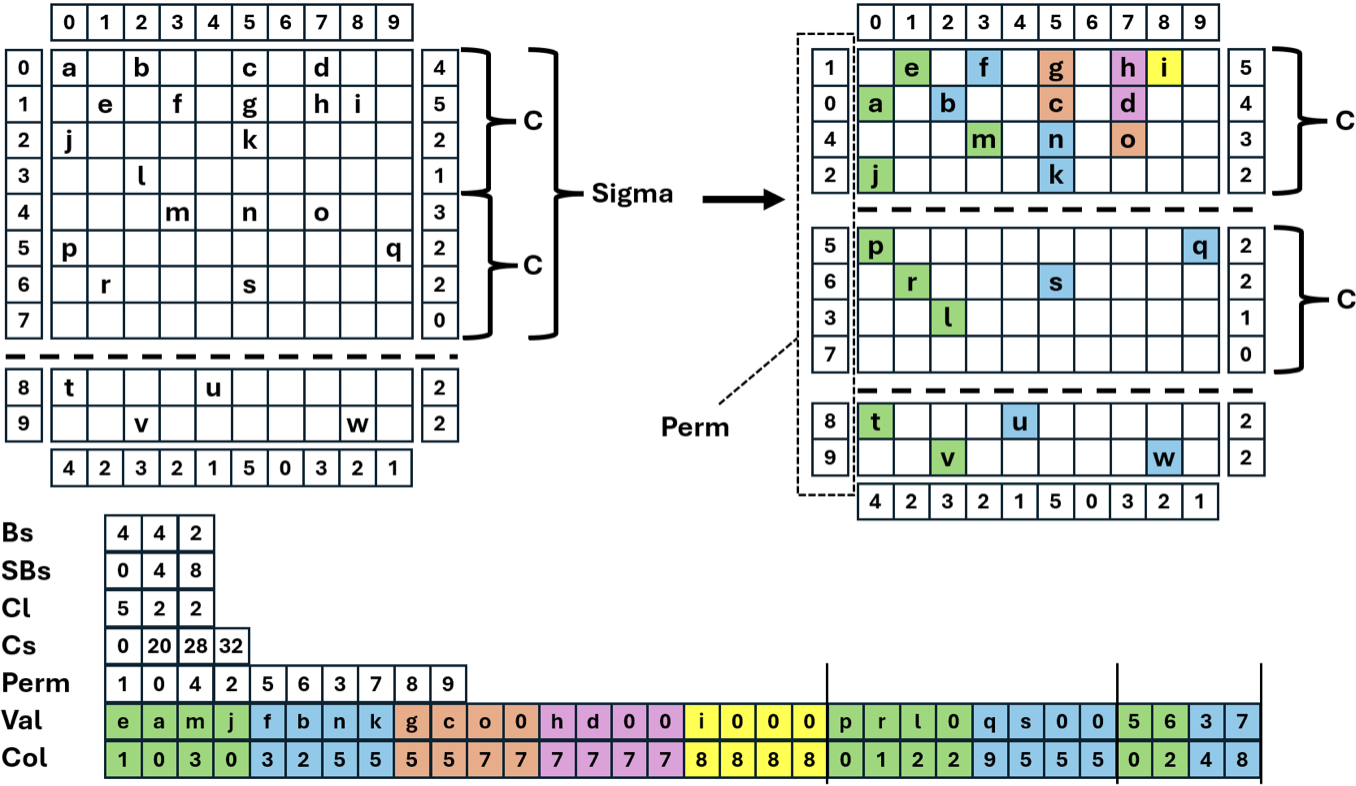}
\caption{Sparse matrix storage scheme using the Sell-C-$\sigma$ format.} \label{scs}
\end{figure*}

The Sell-C-$\sigma$ format was introduced in \cite{6} as an ELLPACK-based storage scheme tailored for CPUs with SIMD support and GPUs. To build a matrix in this format, the original matrix is first partitioned into slices of width $\sigma$, and within each slice, rows are sorted by descending nonzero count. The rows in each slice are then grouped into chunks of $C$ rows, and each chunk is padded with zeros to ensure uniform row length within the chunk. Within every chunk, the nonzero values and their column indices are stored column-wise in the \var{Val} and \var{Col} arrays: i.e., the first nonzero element of each row in the chunk is stored first, followed by the second nonzero element of each row, and so forth (\cref{scs}). In addition, metadata describing chunk sizes and offsets is stored, together with the global row permutation \var{perm}.

The vectorized Sell-C-$\sigma$ SpMV algorithm is shown in \cref{alg:sellcsigma}. The algorithm assumes that the vector register holds \var{bs} double-precision values.

\begin{figure*}[!tb]
\begin{lstlisting} [ caption = {SpMV algorithm for Sell-C-$\sigma$ format}, label = {alg:sellcsigma}]
void SP_MV_SELL_C_SIGMA_RVV(SELL_C_SIGMAmatrix mat, double* b, double* y, double alpha, double beta) {
	#pragma omp parallel for schedule(dynamic, 16)
	for(int i = 0; i < mat.cnt_b; i++) {
		int cur_pos = SBs[i];
		uint32_t bs = __riscv_vsetvlmax_e64m2();
		for (int k = 0; k < Bs[i]; k+= bs) {
			bs = __riscv_vsetvl_e64m2(Bs[i] - k);
			vfloat64m2_t v_py = __riscv_vfmv_v_f_f64m2(0.0, bs);
			for (int j = 0; j < Cl[i]; j++) {
				vuint32m1_t index = __riscv_vle32_v_u32m1(Col + Cs[i]+ j * Bs[i] + k, bs); /*@\label{line:scs1}@*/
				vuint32m1_t index_shift = __riscv_vsll_vx_u32m1(index, 3, bs);
				vfloat64m2_t v_val = __riscv_vle64_v_f64m2(Val + Cs[i] + j * Bs[i] + k, bs);
				vfloat64m2_t v_b = __riscv_vloxei32_v_f64m2(b, index_shift, bs);
				v_py = __riscv_vfmacc_vv_f64m2(v_py, v_val, v_b, bs); /*@\label{line:scs2}@*/
			}
			vuint32m1_t index_perm = __riscv_vle32_v_u32m1(Perm + cur_pos + k, bs); /*@\label{line:scs3}@*/
			vuint32m1_t index_perm_shift = __riscv_vsll_vx_u32m1(index_perm, 3, bs); 
			vfloat64m2_t v_py_c = __riscv_vloxei32_v_f64m2(y, index_perm_shift, bs);
			v_py_c = __riscv_vfmul_vf_f64m2(v_py_c, beta, bs);
			v_py = __riscv_vfmadd_vf_f64m2(v_py, alpha, v_py_c, bs);
			__riscv_vsoxei32_v_f64m2(y, index_perm_shift, v_py, bs); /*@\label{line:scs4}@*/
		}
	}
}
\end{lstlisting}
\end{figure*}

In the pseudo-code, computations are parallelized block-wise. For each block, the vector register \var{v\_py} accumulates partial sums of the products of $C$ rows of the matrix and vector \var{b}. This is done by multiplying each block of \var{bs} nonzeros by their corresponding elements in vector \var{b} (lines \ref{line:scs1}--\ref{line:scs2}). After processing the block, the updated \var{y} values are loaded and stored using the permutation computed during the Sell-C-$\sigma$ conversion (lines \ref{line:scs3}--\ref{line:scs4}).

Notably, the row permutation and column-wise block layout enable the use of contiguous loads rather than indexed loads for the matrix entries, while also eliminating expensive reductions. Furthermore, the vector length parameter remains fixed for most of the intrinsics used. For certain matrices, these factors can provide considerable performance gains over the vectorized CSR version.

\subsection{CSR5 Format}
The CSR5 format was introduced in \cite{8} as an extension of CSR for CPUs with SIMD units and GPUs. It offers good load balancing under parallel execution and supports uniform storage of matrices with arbitrary nonzero patterns.

\begin{figure*}[!tb]
\centering
\includegraphics[width=0.8\textwidth]{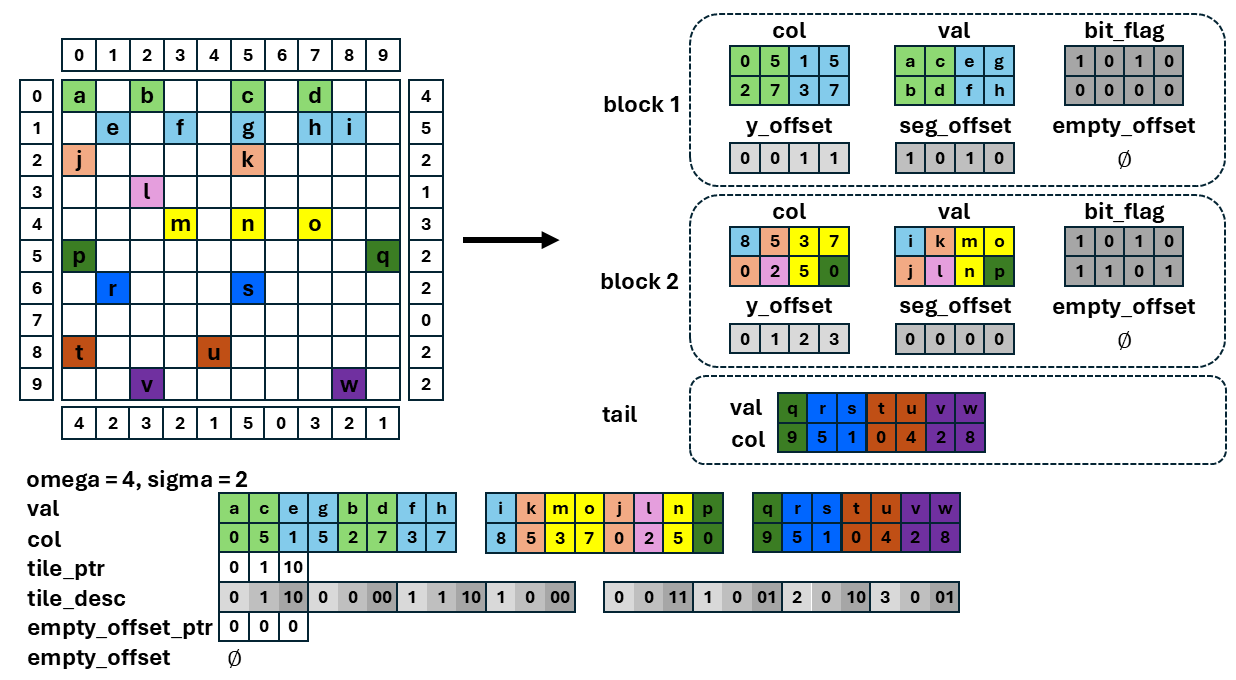}
\caption{Sparse matrix storage scheme using the CSR5 format. It is assumed that tiles contain no empty rows.} \label{csr5}
\end{figure*}

In the CSR5 format, nonzeros and their column indices are stored in fixed-size dense 2D tiles (\cref{csr5}). The format has two parameters, $\omega$ and $\sigma$. $\omega$ is the tile width, which corresponds to the vector register length, and $\sigma$ is the tile height, which is a tunable parameter chosen experimentally. Elements within a tile are stored column-wise so that a single SIMD lane processes an entire column during SpMV. The position of a tile in the matrix is determined by the row number of its first element, which is stored in the \var{tileptr} array; empty rows within a tile are also marked in \var{tileptr}. In \cref{csr5}, such markers are shown as negative values. Additionally, each tile maintains a descriptor that encodes its internal structure. Let $B$ be the tile matrix and $j$ be the current column index. For each column, the tile descriptor includes the following:

\begin{itemize}[itemsep=2pt, parsep=0pt, topsep=2pt, partopsep=0pt]
    \item \var{bit\_flag} -- an array of bits of length $\sigma$, with \var{bit\_flag[i] = 1} if the corresponding entry $B[i][j]$ is the first nonzero in its row. For the correctness of the SpMV algorithm, the first element of the first column of the tile is also marked in the \var{bit\_flag} array;
    \item \var{y\_offset} -- an integer array used to compute the index in vector $y$ where the product should be written;
    \item \var{seg\_offset} -- an integer array used for rows that occupy multiple columns within the tile. It specifies the number of subsequent columns that are completely filled by the last row beginning in the current column;
    \item \var{empty\_offset} -- an array shared across all columns, utilized when the tile contains empty rows to maintain correct indexing into vector $y$.
\end{itemize}

In our implementation, \var{bit\_flag}, \var{y\_offset}, and \linebreak \var{seg\_offset} are stored as bit fields within one or more 32-bit integers in the \var{tiledesc} array. The \var{empty\_offset} entries for all blocks are stored in a single shared array.

The SpMV operation for a matrix in CSR5 format can be performed independently per tile. When all entries of a tile belong to the same row of the original matrix, no cross-column accumulation is required, and the product is computed in the same way as in CSR. The pseudo-code for this case is given below (\cref{alg:csr5-1}). The computation proceeds row by row through the tile. As in CSR, for each row the corresponding values from the \var{Val} and \var{b} arrays are loaded, an FMA operation is performed, and the products are accumulated into the vector register \var{sum}.

If a block contains elements from two or more rows of the original matrix, a more complex algorithm is used. Within each column of a tile, the elements belonging to the same matrix row form contiguous segments. These segments can be classified into three categories: \enquote{red} -- the row is the first in the tile or starts in one of the previous columns; \enquote{green} -- the row lies entirely within the column; and \enquote{blue} -- the row extends beyond the current tile. When a tile spans multiple rows, the partial results must be combined appropriately. The pseudo-code for this case is shown in \cref{alg:csr5-2}. 
During row processing, whenever a \enquote{green} segment ends in a column, the corresponding entry in the \var{sum} register holds the complete result. This accumulated value is stored in vector \var{y} (lines \ref{line:csr5_14}--\ref{line:csr5_20}). All such columns are marked in a mask vector register \var{green\_mask}.

If a \enquote{red} segment ends, the result is stored in \var{first_sum} for later use (lines \ref{line:csr5_21}--\ref{line:csr5_27}). All such columns are marked in \var{red_mask}. After the loop terminates, \var{sum} holds the results for the \enquote{blue} segments marked in \var{blue_mask}. These need to be combined with the values in \var{first_sum} to compute the correct result for rows spanning multiple tile columns. To do so, \var{sum} is copied to \var{last_sum}; \var{first_sum} is copied to \var{sum} and shifted left by one lane; a prefix sum is computed on \var{sum}, and the resulting values are added to the corresponding elements of \var{last_sum} according to the blue segment mask (lines \ref{line:csr5_28}--\ref{line:csr5_35}). As a result, \var{last_sum} contains the correct value for rows with \enquote{blue} segments, which is then stored in \var{y} (\cref{line:csr5_36}).

In each tile, the first row of the original matrix is treated in a special way due to parallel tile processing. In our implementation, static OpenMP scheduling is used to distribute tiles among threads. Consequently, each thread computes a product for a contiguous segment of rows of the original matrix. Under this scheme, the value computed for the first row in the row segment may be incorrect, because part of the product for that row may be computed by the preceding thread. For such rows, results from different threads must be combined correctly. This is done as follows. A shared array \var{calibrator} of length equal to the thread count is used. Each thread stores its portion of the product for the first row it processes in this array (lines \ref{line:csr5_37}--\ref{line:csr5_40}). After the parallel region, the entries of the \var{calibrator} array are used to update vector \var{y} at the positions corresponding to the conflicting rows.

\begin{figure*}[!tbp]
\begin{lstlisting} [caption = {SpMV algorithm for CSR5 format for tile containing elements from one row of the matrix}, label = {alg:csr5-1}]
void SP_MV_CSR5_RVV_SIMPLE_LOOP(CSR5matrix mat, FLOAT* b, FLOAT* y, FLOAT alpha) {
	size_t vl = __riscv_vsetvlmax_e64m2();
	vfloat64m2_t sum = __riscv_vfmv_v_f_f64m2(0.0, vl);
	for (int i = 0; i < sigma; i++) { /*@\label{line:csr5_simple_4}@*/
		vuint32m2_t vcol = __riscv_vle32_v_u32m2(mat.col + i * omega, vl);
		vcol = __riscv_vsll_vx_u32m2(vcol, 3, vl);
		vfloat64m2_t vx = __riscv_vloxei32_v_f64m2(b, vcol, vl);
		vfloat64m2_t vval = __riscv_vle64_v_f64m2(mat.val + i * omega, vl);
		sum = __riscv_vfmadd_vv_f64m2(vval, vx, sum, vl);
	} /*@\label{line:csr5_simple_10}@*/
	vfloat64m2_t vzero = __riscv_vfmv_v_f_f64m2(0.0, vl); /*@\label{line:csr5_simple_11}@*/
	sum = __riscv_vfredusum_vs_f64m2_f64m1(sum, vzero, vl);
	vfloat64m1_t res = __riscv_vfmv_f_s_f64m1_f64(sum, vl); /*@\label{line:csr5_simple_13}@*/
	if (thread_first_row) /*@\label{line:csr5_simple_14}@*/
		mat.calibrator[thread_id] += res * alpha;
	else
		y[current_row] += res * alpha; /*@\label{line:csr5_simple_17}@*/
}
\end{lstlisting}
\begin{lstlisting} [caption = {SpMV algorithm for CSR5 format for tile containing elements from different rows of the matrix}, label = {alg:csr5-2}]
void SP_MV_CSR5_RVV(CSR5matrix mat, FLOAT* b, FLOAT* y, FLOAT alpha) {
	size_t vl = __riscv_vsetvlmax_e64m2();
	vfloat64m2_t sum = __riscv_vfmv_v_f_f64m2(0.0, vl);
	vfloat64m2_t first_sum = __riscv_vfmv_v_f_f64m2(0.0, vl);
	vfloat64m2_t vval = __riscv_vle64_v_f64m2(mat.val, vl);
	vuint32m2_t vcol = __riscv_vle32_v_u32m2(mat.col, vl);
	vcol = __riscv_vsll_vx_u32m2(vcol, 3, vl);
	vfloat64m2_t vx = __riscv_vloxei32_v_f64m2(x, vcol, vl);
	sum = __riscv_vfmadd_vv_f64m2(vval, vx, sum, vl);
	for (int i = 1; i < sigma; ++i) {
		vcol = __riscv_vle32_v_u32m2(mat.col + i * vl, vl);
		vcol = __riscv_vsll_vx_u32m2(vcol, 3, vl);
		vx = __riscv_vloxei32_v_f64m2(x, vcol, vl);
		if (__riscv_vcpop_m_b32(green_mask, vl) > 0) { /*@\label{line:csr5_14}@*/
			vfloat64m2_t vy = __riscv_vloxei32_v_f64m2_m(green_mask, y, vy_off, vl);
			vy = __riscv_vfmadd_vf_f64m2(vy, sum, alpha, vl);
			__riscv_vsoxei32_v_f64m2_m(green_mask, y, vy_off, vy, vl);
			sum = __riscv_vmerge_vvm_f64m2(sum, 0.0, green_mask, vl);
			vfloat64m2_t vy_off = __riscv_vadd_vx_u64m2_m(green_mask,vy_off, 1,vl);
		} /*@\label{line:csr5_20}@*/
		if (__riscv_vcpop_m_b32(red_mask, vl) > 0) { /*@\label{line:csr5_21}@*/
			first_sum = __riscv_vmerge_vvm_f64m2(first_sum, sum, red_mask, vl);
			sum = __riscv_vmerge_vvm_f64m2(sum, 0.0, red_mask, vl);
		}
		vval = __riscv_vle64_v_f64m2(val + i * vl, vl);
		sum = __riscv_vfmadd_vv_f64m2(vval, vx, sum, vl);
	} /*@\label{line:csr5_27}@*/
	vfloat64m2_t last_sum = __riscv_vse64_v_f64m2(sum, vl); /*@\label{line:csr5_28}@*/
	sum = __riscv_vmerge_vvm_f64m2(sum, first_sum, first_mask, vl);
	sum = __riscv_vslidedown_vx_f64m2(sum, 1, vl);
	vfloat64m2_t tmp_sum = __riscv_vse64_v_f64m2(sum, vl);
	sum = prefix_sum(sum);
	vfloat64m2_t sum_perm = __riscv_vrgather_vv_f64m2(sum, perm, vl);
	sum = __riscv_vfadd_vv_f64m2(   __riscv_vfsub_vv_f64m2(sum_perm, sum), tmp_sum);
	last_sum = __riscv_vfadd_vv_f64m2(blue_mask, last_sum, sum, vl); /*@\label{line:csr5_35}@*/
	__riscv_vsoxei32_v_f64m2_m(blue_mask, y, vy_off,last_sum, vl); /*@\label{line:csr5_36}@*/
	if (thread_first_row) /*@\label{line:csr5_37}@*/
		mat.calibrator[thread_id] += 
        first_prod * alpha;
	else
		y[first_row] += first_prod * alpha; /*@\label{line:csr5_40}@*/
}
\end{lstlisting}
\end{figure*}

The implementations of the SpMV algorithm in CSR5 format using AVX instructions and RVV instructions are generally similar. An important difference is that RVV contains more specific instructions, such as vector slide operations, bitwise operations with mask vector registers, and masked instructions. The latter also takes a mask register as input and performs the operation only on those elements for which the corresponding bit in the mask register is set. All these features enable the implementation of the algorithm using fewer instructions. Furthermore, RVV instructions can operate on a group of registers, where the number of registers is determined by the LMUL parameter in the vector configuration. The LMUL affects the number of tile columns and needs to be chosen carefully to achieve the best performance.

\subsection{HCSR Format}
We propose a novel storage format, Hierarchical CSR (HCSR), a block-based variant of CSR. Although block variants of the CSR format have been widely discussed in the literature, we propose to store the block partitioning structure of the original matrix also as a sparse matrix.

In HCSR, we split the matrix into blocks of fixed size $R~\times~C$. Each nonempty block is stored either in CSR or in COO format, depending on its nonzero occupancy; otherwise, it is marked as empty. Note that when the matrix dimensions are not divisible by $R$ and $C$, the last block column contains blocks of size $R \times (M \bmod C)$, and the last block row contains blocks of size $(N \bmod R) \times C$.

We use a switching threshold $p_{\text{fix}}$, computed as the average number of nonzeros per row in the original matrix. The occupancy of a block is defined as $p_{\text{block}} = nz_{\text{block}} / R$, where $nz_{\text{block}}$ is the number of nonzero entries in the block. Based on this threshold, a block is stored in CSR if $p_{\text{block} } > p_{\text{fix}}$; otherwise, COO is used. Since the block decomposition of the matrix, including empty blocks, forms a sparse structure, we store the nonempty block information in CSR format. 

\begin{figure*}[!tb]
\centering
\includegraphics[width=0.8\textwidth]{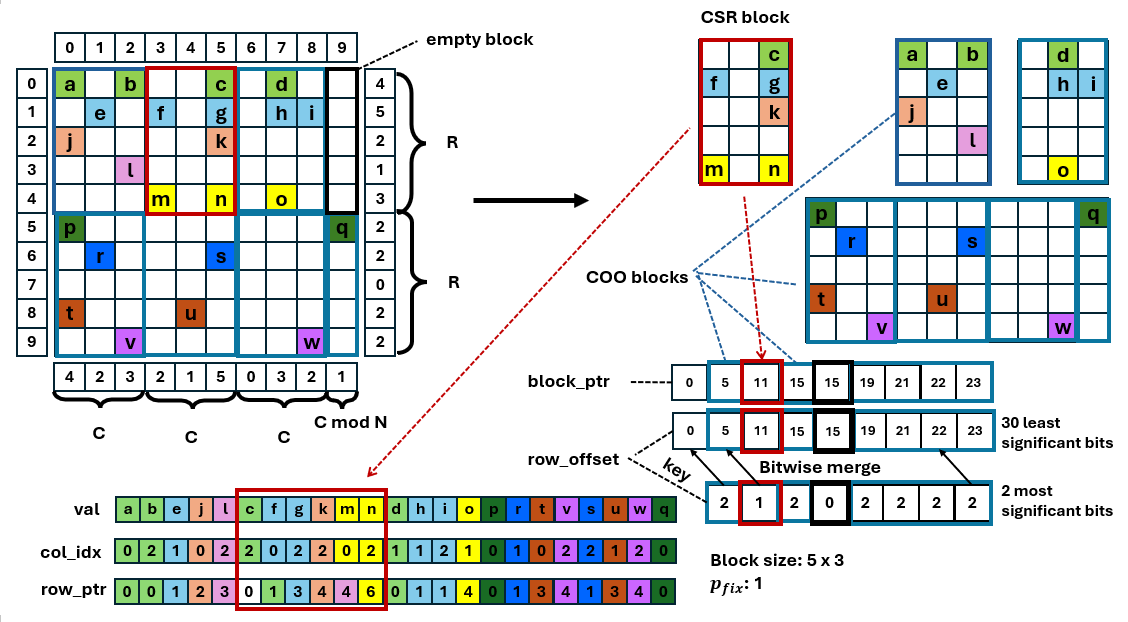}
\caption{Sparse matrix storage scheme using the HCSR format.}
\label{CSRvalentin}
\end{figure*}

Thus, the HCSR representation of a sparse matrix consists of five arrays (\cref{CSRvalentin}):
\begin{itemize}[itemsep=2pt, parsep=0pt, topsep=2pt, partopsep=0pt]
    \item \var{values} -- an array of nonzero values of the matrix, shared across all blocks;
    \item \var{col\_idx} -- an array of column indices of nonzero values, shared across all blocks;
    \item \var{row\_ptr} -- an array of row-start indices in the \var{col\_idx} array for CSR blocks, or row indices of nonzeros for COO blocks, shared across all blocks;
    \item \var{row\_offset} -- an array where the lower 30 bits of each element store the offset into the \var{row\_ptr} array for each block, and the upper 2 bits store a \var{key} value describing the block type (0 for empty, 1 for CSR, or 2 for COO);
    \item \var{block\_ptr} -- an array of offsets into the \var{col\_idx} and \var{values} arrays for each block.
\end{itemize}

The HCSR-based SpMV algorithm performs a parallel traversal of the matrix, proceeding row-wise through the nonempty block rows (\cref{alg:csr_valentin}). Unlike the standard CSR format, which operates on individual matrix entries and vector elements, the HCSR scheme computes the product of each nonempty block of the matrix with the corresponding segment of vector \var{b}. Compared to traditional CSR, the use of the new format still preserves efficient processing of nonzero values in the matrix, while improving the locality of accesses to the vector \var{b}.

\begin{lstlisting}[float=*, caption={SpMV algorithm for HCSR format}, label={alg:csr_valentin}]
void SPMV_HCSR(HCSRmatrix mat, double* b, double* y, double alpha, double beta) {
	#pragma omp parallel for schedule(dynamic)
	for (int i = 0; i < mat.m; i += mat.R) {
		int maxr = min(mat.R, mat.m - i);
		for (int r = 0; r < maxr; ++r) {
			y[i + r] *= beta;
			for (int j = 0; j < mat.n; j += mat.C) {
				int block = j / mat.C + (i / mat.R) * mat.b_n; /*@\label{line:hcrs1}@*/
				int bptr = mat.block_ptr[block];
				int row_ptr_offset = mat.row_offset[block]; 
				int mask = 0x3FFFFFFF;
				int key = row_ptr_offset >> 30;
				row_ptr_offset &= mask; /*@\label{line:hcrs6}@*/
				int nz = mat.block_ptr[block + 1] - bptr; /*@\label{line:hcrs2}@*/
				if (key == 1) 
					KERNEL_HCSR_CSR(alpha, mat.row_ptr + row_ptr_offset, mat.col_idx + bptr, mat.values + bptr, b + j, y + i, maxr); /*@\label{line:hcrs3}@*/
				else 
					KERNEL_HCSR_COO(alpha, mat.row_ptr + row_ptr_offset, mat.col_idx + bptr, mat.values + bptr, b + j, y + i, nz); /*@\label{line:hcrs4}@*/
			} 
		}
	}
}

void KERNEL_HCSR_COO(double alpha, int* row_ptr, int* col_idx, double* values, double* b, double* y, int nz) {
	for (int i = 0; i < nz; ++i)
		y[row_ptr[i]] += alpha * values[i] * b[col_idx[i]];
}
\end{lstlisting}

In the pseudocode, for each block being processed (lines \ref{line:hcrs1}--\ref{line:hcrs4}), we first compute its parameters: the block number \var{block} (line \ref{line:hcrs1}), the number of nonzeros $nz$  (line \ref{line:hcrs2}), the offset \linebreak \var{row\_ptr\_offset} into the \var{row\_ptr} array (line \ref{line:hcrs6}), and the block type \var{key}. Depending on the block type, either \linebreak \var{KERNEL\_HCSR\_CSR} (line \ref{line:hcrs3}) or \var{KERNEL\_HCSR\_COO} (line \ref{line:hcrs4}) is called. The \var{KERNEL\_HCSR\_CSR} function is essentially the same as the vectorized CSR SpMV implementation in \cref{alg:CSR}, aside from the vector register length and the multiplication of $y$ by $\beta$. For \var{KERNEL\_HCSR\_COO}, we use a scalar implementation, as experimental evaluation demonstrated that the vectorized version of this function is less performant than the scalar one.

Thus, we obtain the two-level matrix storage scheme that combines the advantages of standard sparse formats and, in many cases, improves cache efficiency. Note that the sparse block representation approach can be extended recursively, for example, during distributed processing of huge sparse matrices. For this reason, we call the new format hierarchical, although in this paper we limit ourselves to two levels, which are sufficient for the RISC-V processors we use.

\section{Experimental Results}
\label{sectionExp}
\subsection{Experimental setup}

Experiments were carried out on two devices.
The first test system, the Banana Pi BPI-F3 board, features a \textbf{SpacemiT Keystone K1} processor (8x1.6GHz SpacemiT x60 cores with 8-stage in-order dual-issue pipeline, RVA22 Profile and 256-bit RVV 1.0 standard) with 16 GB of RAM and the Bianbu 2.1 operating system. The Banana Pi BPI-F3 (8x X60) has a theoretical peak double precision performance of 102.4 GFLOPs and a single precision performance of 204.8 GFLOPs. 

The second test system is based on the \textbf{SpacemiT Keystone K3} processor (8x SpacemiT X100 high-performance cores clocked at up to 2.4 GHz with 12-stage out-of-order quad-issue pipeline, RVA23 Profile and RVV 1.0 standard with VLEN=256 and 8x SpacemiT A100 AI cores clocked at up to 2.1 GHz with 8-stage in-order dual-issue pipeline, RVA22 Profile and RVV 1.0 standard with VLEN=1024) with 8 GB of RAM \linebreak (LPDDR5-6400) and the Bianbu 4.0.1 operating system. The SpacemiT K3 high-performance cores (8x X100) feature a theoretical peak double precision performance of 153.6 GFLOPs and a single precision performance of 307.2 GFLOPs. The SpacemiT K3 AI cores (8x A100) offer a theoretical peak double precision performance of 134.4 GFLOPs, a single precision performance of 537.6 GFLOPs, and up to 60 TOPS of AI performance. All our experiments were performed using only X100 cores. A GCC RISC-V 14.2.0 cross-compiler was used on both test systems.

\subsection{Baseline selection}
The choice of a baseline against which all other algorithms are evaluated is a critical consideration in any  computational study. In our context, this choice is far from straightforward due to the large configuration space: one can vary storage formats along with their hyperparameters, toggle between scalar and vector implementations with different vector length multipliers (LMUL), and apply sequential or parallel execution models using diverse load-balancing strategies. Furthermore, these variants can be built under various compiler flags, benchmarked across different hardware platforms, and evaluated on a broad matrix collection spanning disparate sparsity structures.

Initially, we considered the vectorized SpMV implementation for the CSR format as our primary baseline. However, we soon observed that it significantly underperformed relative to the naive scalar version on a considerable number of matrices. For this reason, we adopted the scalar CSR SpMV as our baseline for both sequential and parallel evaluations. Nevertheless, the vector CSR implementation remains fully integrated into our experimental evaluation alongside other sparse formats.

\subsection{Benchmark Suite}

A set of 121 matrices from the SuiteSparse Matrix collection \cite{19} was used in the experiments. Our test set covers matrices from a variety of application domains, including HPC applications and graph processing. Many of them have been used in previous SpMV-related studies. The benchmark suite included matrices of sizes from 0.5K to 8.39M rows, with fill-in from $3.1 \times 10^{-7}$\% to $1.4 \times 10^{-2}$\%. Detailed information about the test set is given in Supplementary Table S1. For \cref{SpMV_operation}, the dense vector $b$ was filled with random numbers from the interval $[0, 1)$, and the scalar coefficients $\alpha$ and $\beta$ were fixed to $1.0$ and $0.0$, respectively. For the formats that support tunable block sizes (Sell-C-$\sigma$, CSR5, VHCC, HCSR), we pre-tuned the parameter sets to find the combinations that minimize the computation time on most test matrices, separately for single and double precision. All experiments were run using 1 and 8 cores in both single and double precision. For each format, the minimum time of a single SpMV operation across five independent runs was used for performance comparison. For each test case, we compare the results by taking the ratio of the SpMV computation time for a given matrix and format to the run time of the scalar CSR SpMV implementation. The relative speedups over CSR are presented as heatmaps in \cref{sectionExpRes}. Yellow indicates values close to 1.0 (no speedup), red marks the lowest performance (near $0.0$), and green denotes the best values ($>1.0$). The matrices are sorted by increasing fill-in.

We have observed that the performance of vectorized SpMV implementations is highly sensitive to the choice of the LMUL parameter. For highly sparse matrices, the computation time can differ by a factor of two or three. Thus, to maximize performance, LMUL must be chosen by taking into account both the hardware and the matrix structure. For instance, in the CSR format, the fastest execution for highly sparse matrices was obtained with LMUL=1 in single precision and LMUL=2 in double precision. For denser matrices, LMUL=4 performed best in both precisions. For the results presented below, a single LMUL value was selected per format for all test matrices, with separate tuning for single and double precision.

\subsection{SpMV performance comparison}
\label{sectionExpRes}

\subsubsection{SpacemiT K1}

\begin{figure*}[!tb]
    \centering
    \includegraphics[width=\textwidth]{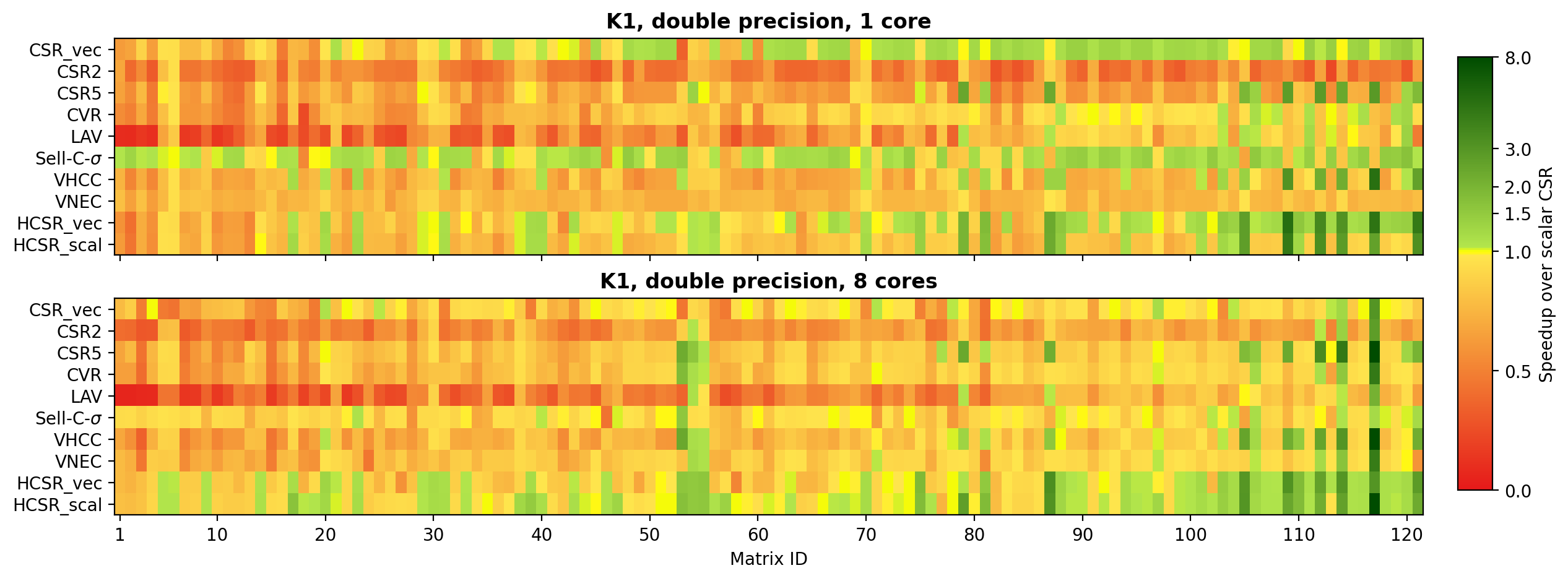}
    \vspace{0.1cm}
    \includegraphics[width=\textwidth]{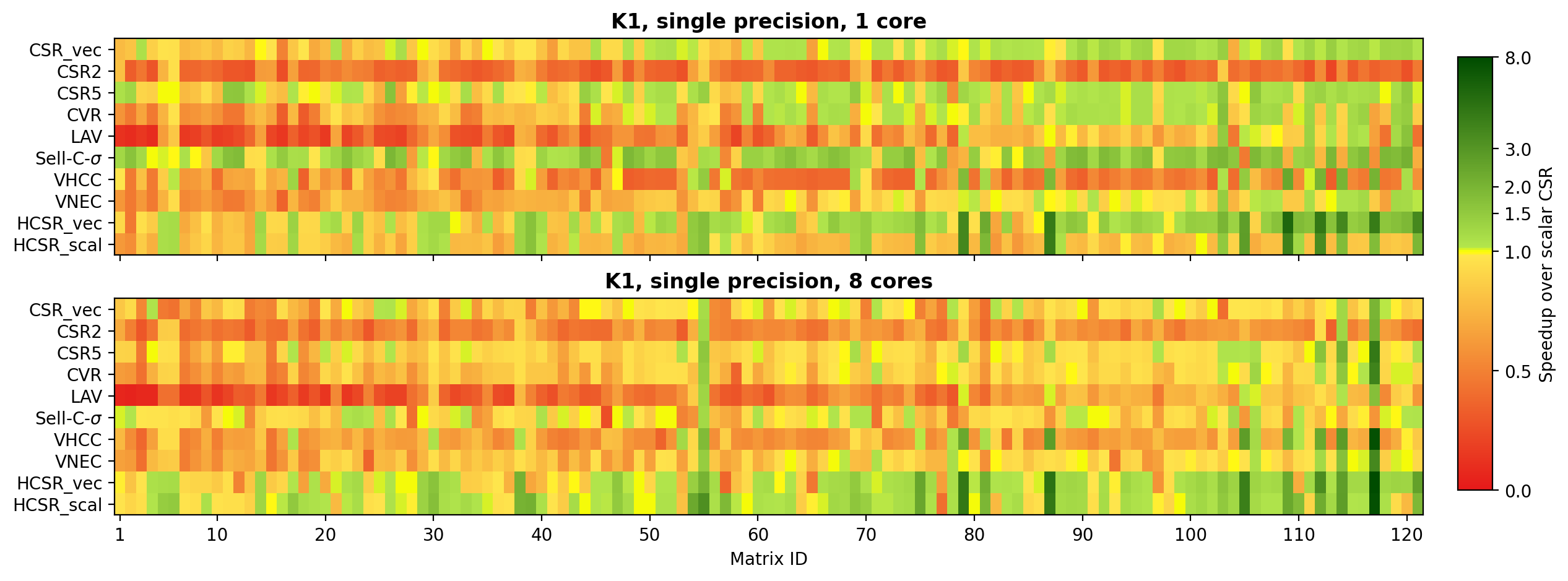}
    \caption{SpMV speedup for various sparse formats relative to scalar CSR on SpacemiT K1.}
    \label{fig-combined-heatmap-k1}
\end{figure*}

\begin{figure*}[!tb]
    \centering
    \includegraphics[width=\textwidth]{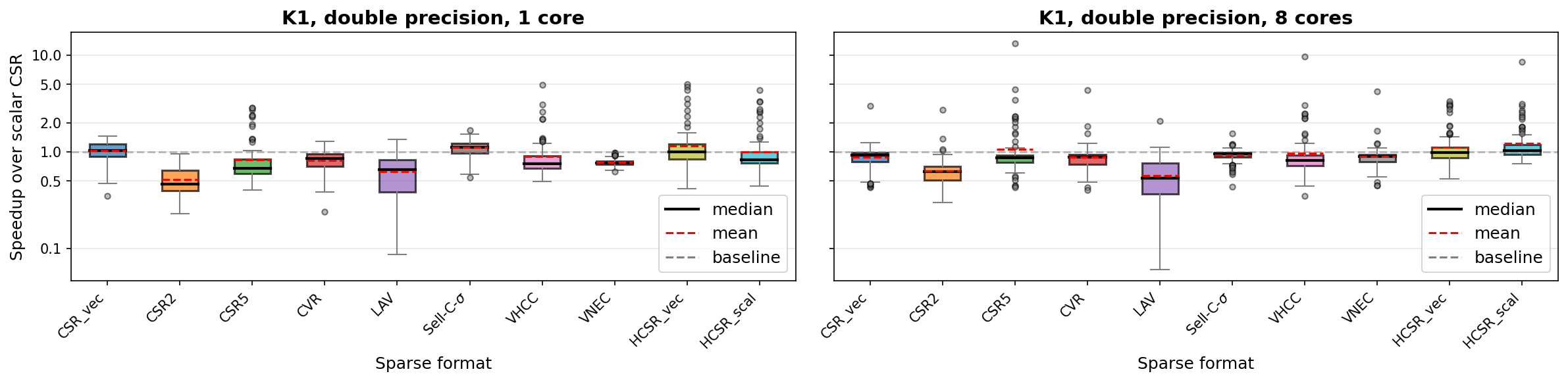}
    \vspace{0.1cm}
    \includegraphics[width=\textwidth]{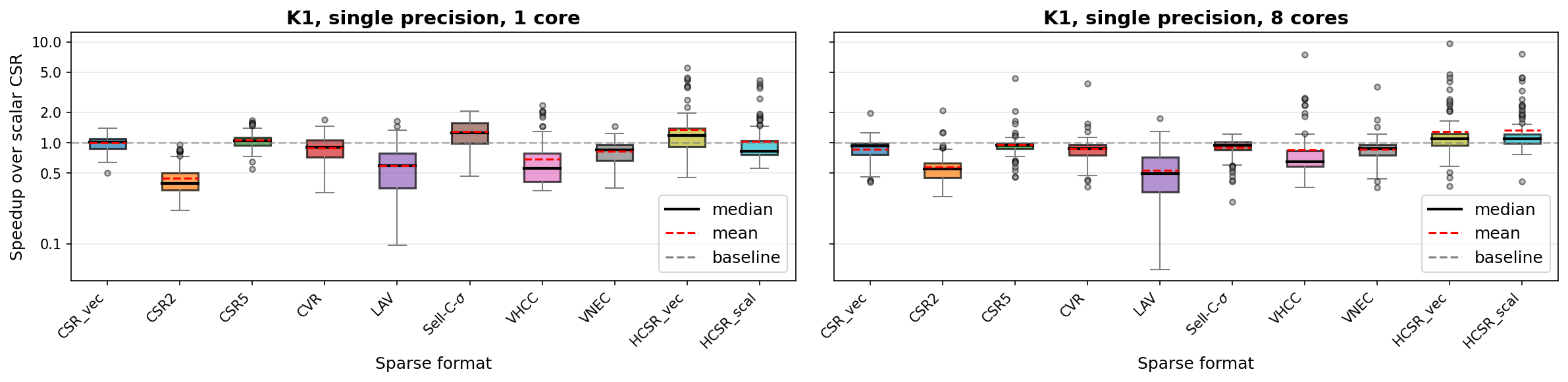}
    \caption{Distribution of SpMV speedup for various sparse formats relative to scalar CSR on SpacemiT K1.}
    \label{fig-combined-boxplot-k1}
\end{figure*}

For the SpacemiT K1 platform, \cref{fig-combined-heatmap-k1} compares the execution time of a single SpMV operation for all implemented formats relative to the scalar CSR baseline. \Cref{fig-combined-boxplot-k1} presents the corresponding speedup distributions.

Based on these results, we summarize the performance of each sparse format as follows:
\begin{itemize}[itemsep=2pt, parsep=0pt, topsep=2pt, partopsep=0pt]
\item For denser matrices, the \textbf{vectorized CSR} implementation is faster than its scalar counterpart on a single core and performs similarly on 8 cores. It outperforms the baseline on half of the test matrices, delivering an average speedup of 19\% in double precision and 13\% in single precision for this subset. On sparser matrices, the vectorized implementation is 17\% slower on average.
\item \textbf{Sell-C-$\sigma$} achieves the best performance among all evaluated formats in single-core execution for both single and double precision. When it outperforms scalar CSR, it achieves average speedups of 21\% in double precision and 46\% in single precision. On 8 cores, however, its computation time is slower than or comparable to that of scalar CSR.
\item On a single core, both \textbf{vectorized} and \textbf{scalar HCSR} outperform scalar CSR on most test matrices. When outperforming the baseline, they demonstrate an average speedup of 1.5$\times$ in double precision and 1.6$\times$ in single precision. On 8 cores, both HCSR implementations continue to outperform scalar CSR; however, on half of the test matrices, which are sparser, the scalar version runs faster than the vectorized one. In these successful cases, scalar and vectorized HCSR run 1.4$\times$ faster in double precision and 1.5$\times$ faster in single precision on average. The scalar HCSR implementation in double precision and both HCSR implementations in single precision exhibit a large variance in speedup relative to CSR. The largest speedups were obtained on Kronecker and highly rectangular matrices: up to 5$\times$ on a single core in double precision and 3.5$\times$ in single precision, and up to 9.6$\times$ on 8 cores.
\item \textbf{CSR5} performs worse than scalar CSR in double precision, but significantly outperforms it on 20 rectangular and denser matrices, with average speedups of 1.8$\times$ in single-core and 2.7$\times$ in multi-core execution. In single precision, CSR5 is up to 1.8$\times$ faster than CSR in single-core execution (outperforming the baseline on two-thirds of the test matrices) and is comparable to CSR in multi-core execution.
\item \textbf{VHCC} performs similarly to scalar CSR in double precision and worse in single precision on most matrices. Similar to CSR5, it outperforms CSR on rectangular and denser matrices, with average speedups of 1.6$\times$ in single-core and 2.0$\times$ in multi-core execution for double precision.
\item \textbf{CVR} and \textbf{VNEC} perform worse than scalar CSR in double precision and similarly in single precision. However, as the number of threads increases, their performance gap narrows, and on 8 cores they outperform CSR on a small number of matrices.
\item \textbf{CSR2} and \textbf{LAV} generally perform worse than CSR, with an average slowdown of 2$\times$. However, on several specific matrices, LAV outperforms scalar CSR by an average of 10\% in double precision and 20\% in single precision.
\end{itemize}

In \cite{40}, the authors reported a vectorized CSR performance of 0.05--1.25 GFlops/s on 8 cores of SpacemiT K1 (in double precision). Our vectorized CSR falls within a similar range: 0.03--1.22 GFlops/s, with a mean of 0.67 GFlops/s and variance of 0.15. These results validate the competitiveness of our implementation.

\subsubsection{SpacemiT K3}
We next evaluate SpMV performance on the SpacemiT K3 processor. Compared to the K1, the absolute SpMV execution time is reduced on average by 3$\times$ in double precision and 2$\times$ in single precision. For web graphs, Kronecker matrices, and some rectangular matrices, the reduction reaches up to 18$\times$ on a single core for selected formats.

\begin{figure*}[!tb]
    \centering
    \includegraphics[width=\textwidth]{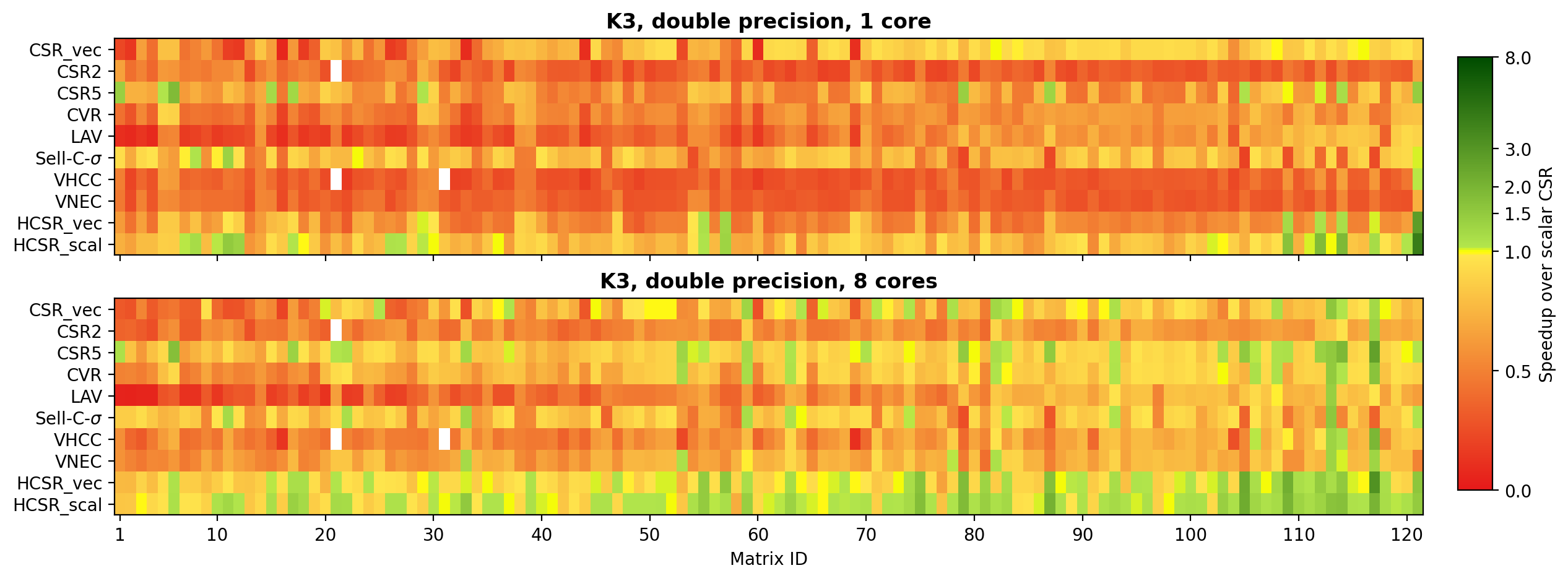}
    \vspace{0.1cm}
    \includegraphics[width=\textwidth]{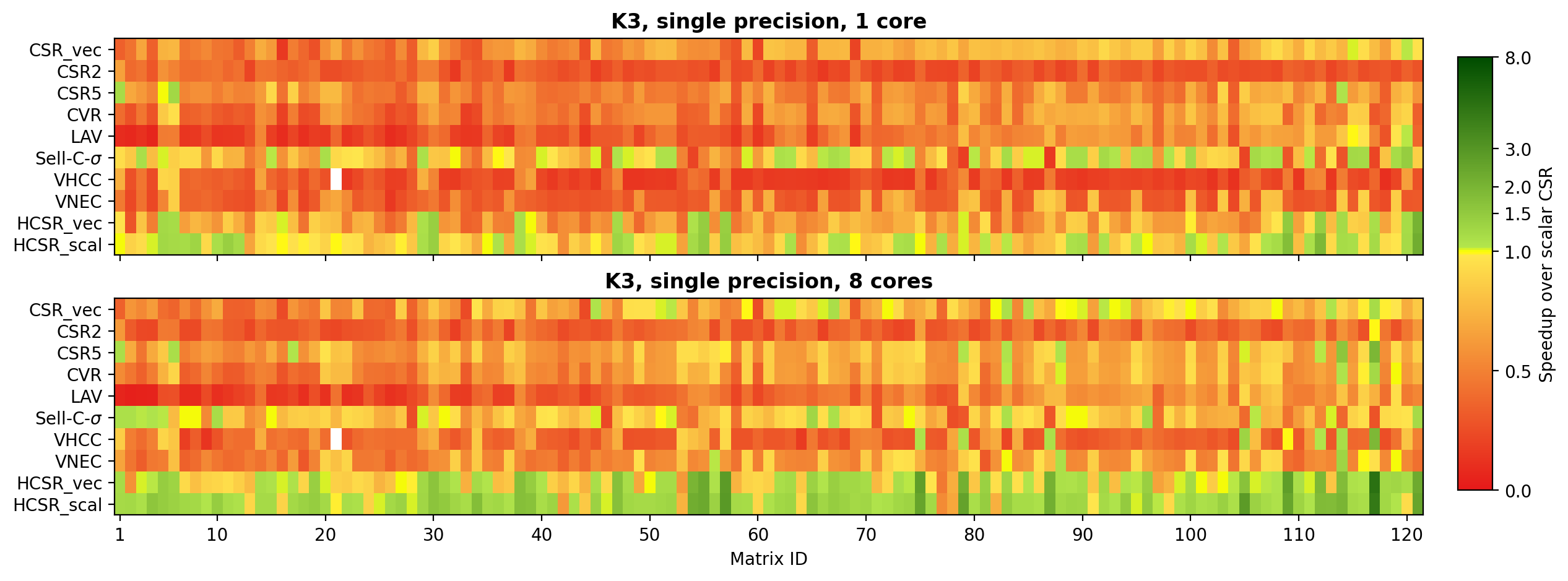}
    \caption{SpMV speedup for various sparse formats relative to scalar CSR on SpacemiT K3.}
    \label{fig-combined-heatmap-k3}
\end{figure*}

\begin{figure*}[!tb]
    \centering
    \includegraphics[width=\textwidth]{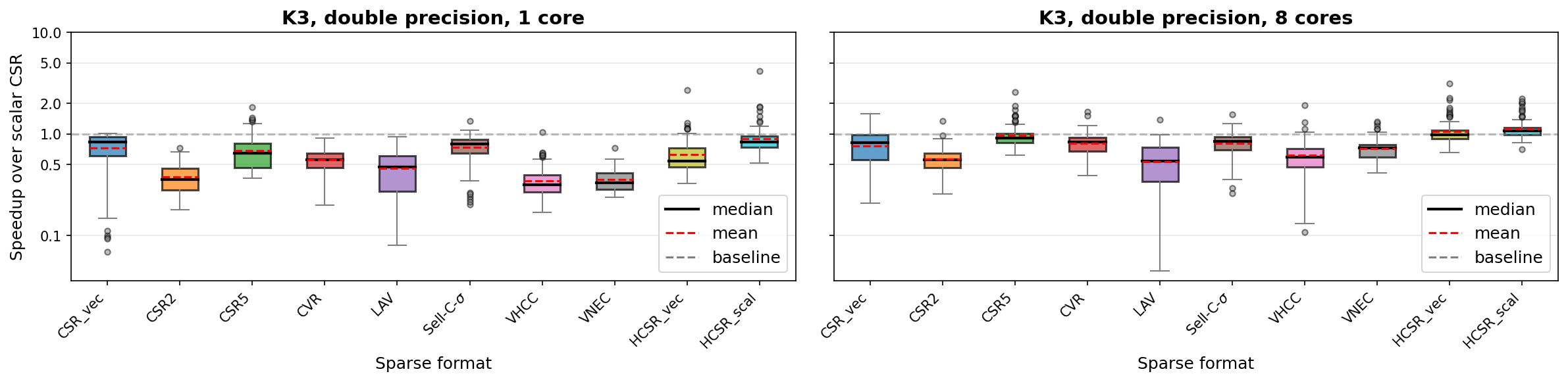}
    \vspace{0.1cm}
    \includegraphics[width=\textwidth]{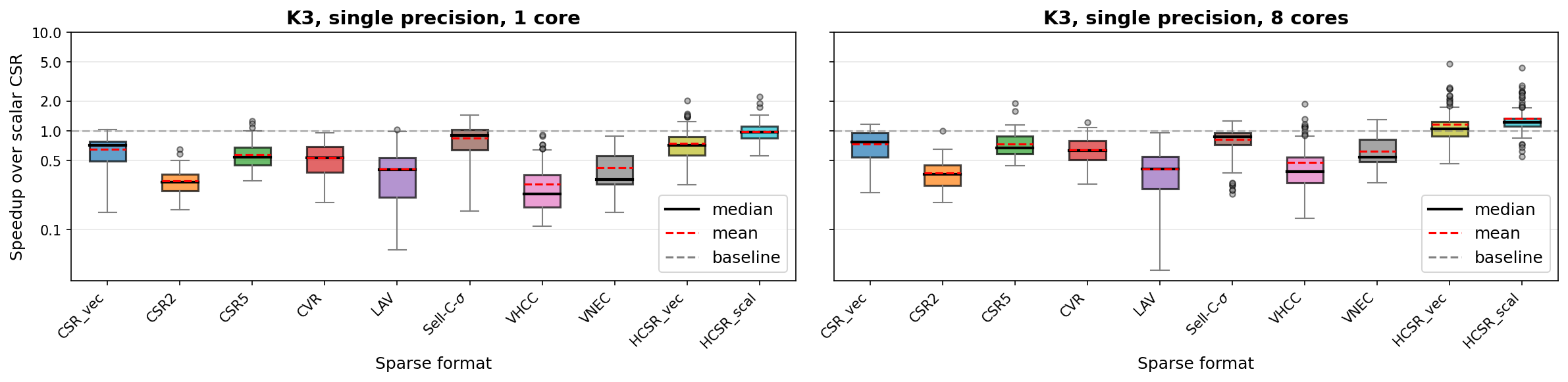}
    \caption{Distribution of SpMV speedup for various sparse formats relative to scalar CSR on SpacemiT K3.}
    \label{fig-combined-boxplot-k3}
\end{figure*}

\begin{figure*}[t!]
    \centering
    \includegraphics[width=0.49\textwidth]{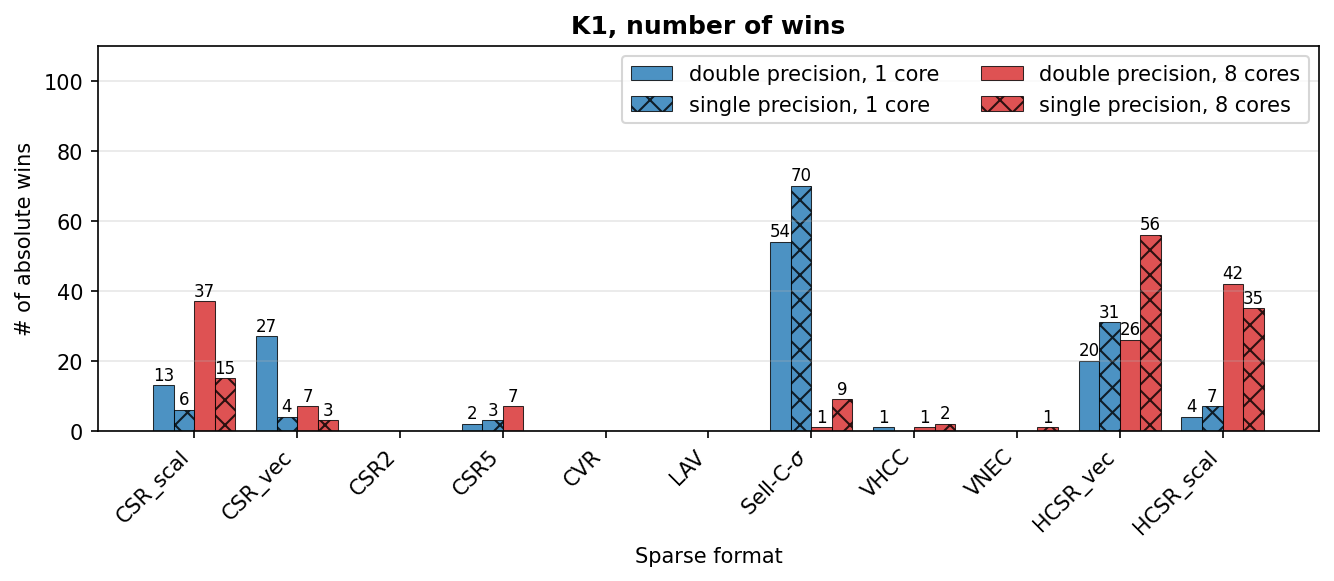}
    \hfill 
    \includegraphics[width=0.49\textwidth]{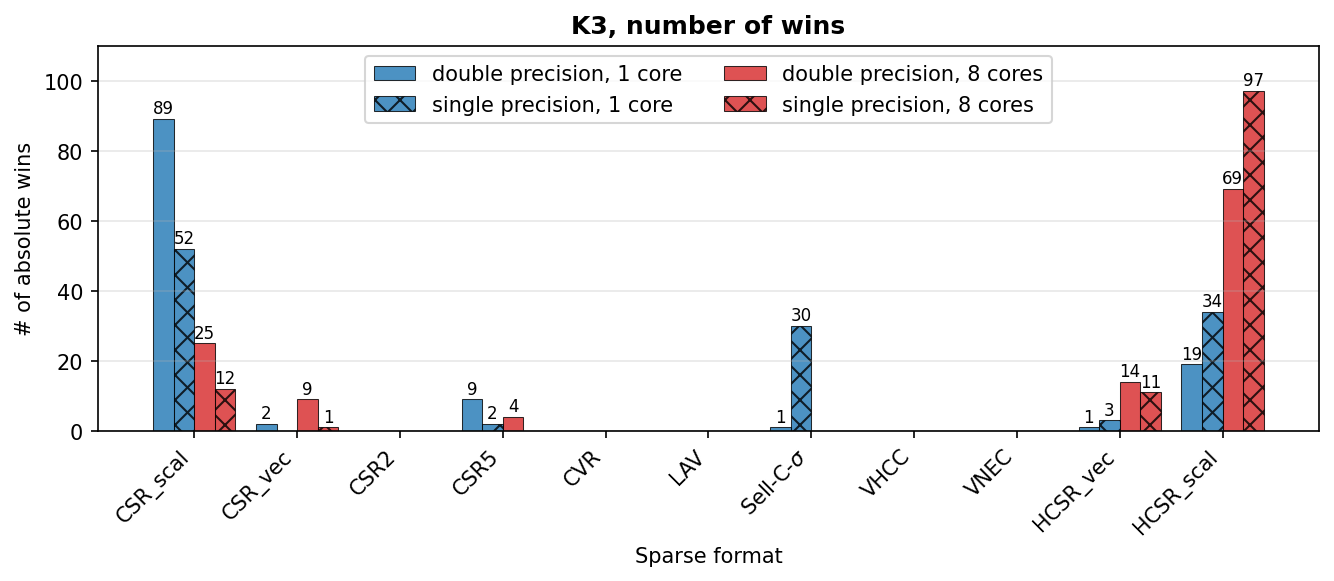}
    
    \caption{Number of matrices where the given format outperforms all others on SpacemiT K1 (left) and SpacemiT K3 (right).}
    \label{fig-wins}
\end{figure*}

\Cref{fig-combined-heatmap-k3} presents the SpMV speedup over scalar CSR for both double and single precision, while \cref{fig-combined-boxplot-k3} illustrates the corresponding distributions. The experimental results yield several key observations:
\begin{enumerate}[itemsep=2pt, parsep=0pt, topsep=2pt, partopsep=0pt]
\item Scalar CSR shows considerably improved performance relative to the other formats, outperforming all other implementations on a substantial portion of the test matrices in single-core execution. The closest results to scalar CSR on a single thread are achieved by HCSR and Sell-C-$\sigma$. On 8 cores, however, HCSR outperforms all other implementations on the majority of matrices (\cref{fig-wins}).
\item On a single core, scalar CSR is faster than vectorized CSR across all test matrices by an average of 30\%. On 8 cores, vectorized CSR outperforms scalar CSR on 14 to 19 matrices, delivering an average speedup of 16\% in double precision and 8\% in single precision.
\item For single precision, Sell-C-$\sigma$ performs comparably to or outperforms CSR on half of the test matrices, achieving an average speedup of 11\%. Similar to the K1 results, the advantage of Sell-C-$\sigma$ decreases as the number of cores increases.
\item Consistent with the trend observed for CSR, scalar HCSR is faster than its vectorized counterpart for almost all matrices, outperforming it by an average of 1.5$\times$ on 1 core and by 18\% on 8 cores. As shown in \cref{fig-wins}, HCSR outperforms all other implementations on the majority of matrices on 8 cores, surpassing CSR by 1.3$\times$ in double precision and 1.4$\times$ in single precision.
\item Among the remaining formats, CSR5 is notable: it outperforms CSR on a quarter of the matrices in double precision on 8 cores (by an average of 25\%), and for several matrices, it surpasses all other implementations.
\end{enumerate}

\subsection{Architectural Impact on SpMV Performance: SpacemiT K1 vs. K3}
\label{sectionAnalysis}
The fast-paced evolution of RISC-V technology drives strong interest in performance evaluation of new hardware, especially when a vendor releases a next-generation device. We therefore analyze how the SpacemiT K3 affects SpMV performance compared to the previous-generation SpacemiT K1. We first highlight the key differences of the new device that are relevant to this study. First, the K3 device features significantly faster memory (LPDDR5-6400 vs. LPDDR4/LPDDR4x-2666). While peak memory bandwidth figures give a rough estimate, the STREAM benchmark provides more meaningful real-world measurements of memory subsystem performance. We adapted our earlier benchmark \cite{49} for RVV~1.0 and used the following experimental methodology. We ran four tests that perform elementary array operations (Copy, Scale, Add, Triad). The array sizes are selected so that the working set fits entirely in L1 cache, L2 cache, and DRAM, respectively. Both scalar and vectorized versions are executed with thread binding aligned to the memory hierarchy. The workload was configured as follows: 1 thread was used for L1 testing (with the result scaled by 8), 4 threads for the shared L2 (scaled by 2), and 8 threads for measuring the global DRAM bandwidth. The results (\cref{tab:stream}) demonstrate a substantial increase in achieved memory bandwidth, from 6.9~GB/s on the K1 to 19.9 GB/s on the K3 -- a speedup of nearly 3$\times$. This improvement is particularly significant for memory-bound algorithms like SpMV, which has an arithmetic intensity of \linebreak $\sim$0.1~FLOP/byte.

\begin{table*}[tb]
\centering
\caption{STREAM benchmark results for SpacemiT K1 (top) and SpacemiT K3 (bottom).}
\label{tab:stream}
\begin{tabularx}{\textwidth}{l *{6}{R} c *{6}{R}}
\toprule
& \multicolumn{6}{c}{\textbf{K1 MB/s}} & & \multicolumn{6}{c}{\textbf{K1 MB/s $\times$ 8 cores}} \\
\cmidrule(lr){2-7} \cmidrule(lr){9-14} 
& \multicolumn{3}{c}{Scalar} & \multicolumn{3}{c}{RVV} & & \multicolumn{3}{c}{Scalar} & \multicolumn{3}{c}{RVV} \\
& \multicolumn{1}{c}{\textbf{L1}} & \multicolumn{1}{c}{\textbf{L2}} & \multicolumn{1}{c}{\textbf{DRAM}} & \multicolumn{1}{c}{\textbf{L1}} & \multicolumn{1}{c}{\textbf{L2}} & \multicolumn{1}{c}{\textbf{DRAM}} & & \multicolumn{1}{c}{\textbf{L1}} & \multicolumn{1}{c}{\textbf{L2}} & \multicolumn{1}{c}{\textbf{DRAM}} & \multicolumn{1}{c}{\textbf{L1}} & \multicolumn{1}{c}{\textbf{L2}} & \multicolumn{1}{c}{\textbf{DRAM}} \\
\midrule
Copy  & 5121  & 27906 & 5752  & 4846  & 27351 & 5832  & & 40966 & 55813 & 5752  & 38770 & 54702 & 5832 \\
Scale & 2432  & 10014 & 6824  & 4909  & 27906 & 5789  & & 19454 & 20028 & 6824  & 39268 & 55813 & 5789 \\
Add   & 3965  & 13836 & 6938  & 6782  & 20462 & 5680  & & 31717 & 27672 & 6938  & 54252 & 40925 & 5680 \\
Triad & 3293  & 12056 & 6955  & 6711  & 20261 & 5663  & & 26346 & 24112 & 6955  & 53687 & 40523 & 5663 \\
\midrule
& \multicolumn{6}{c}{\textbf{K3 MB/s}} & & \multicolumn{6}{c}{\textbf{K3 MB/s $\times$ 8 cores}} \\
\cmidrule(lr){2-7} \cmidrule(lr){9-14} 
& \multicolumn{3}{c}{Scalar} & \multicolumn{3}{c}{RVV} & & \multicolumn{3}{c}{Scalar} & \multicolumn{3}{c}{RVV} \\
& \multicolumn{1}{c}{\textbf{L1}} & \multicolumn{1}{c}{\textbf{L2}} & \multicolumn{1}{c}{\textbf{DRAM}} & \multicolumn{1}{c}{\textbf{L1}} & \multicolumn{1}{c}{\textbf{L2}} & \multicolumn{1}{c}{\textbf{DRAM}} & & \multicolumn{1}{c}{\textbf{L1}} & \multicolumn{1}{c}{\textbf{L2}} & \multicolumn{1}{c}{\textbf{DRAM}} & \multicolumn{1}{c}{\textbf{L1}} & \multicolumn{1}{c}{\textbf{L2}} & \multicolumn{1}{c}{\textbf{DRAM}} \\
\midrule
Copy  & 13554 & 53374 & 19290 & 16207 & 49695 & 19200 & & 108433 & 106749 & 19290 & 129659 & 99391 & 19200 \\
Scale & 9874  & 54600 & 19636 & 15339 & 52358 & 19474 & & 78988  & 109200 & 19636 & 122714 & 104715 & 19474 \\
Add   & 14810 & 48687 & 19681 & 21386 & 43689 & 19858 & & 118482 & 97374  & 19681 & 171086 & 87378  & 19858 \\
Triad & 14810 & 48472 & 18686 & 21166 & 45655 & 18729 & & 118482 & 96944  & 18686 & 169330 & 91309  & 18729 \\
\bottomrule
\end{tabularx}
\end{table*}
 
Second, it is worth noting that the K3 employs an Out-of-Order (OoO) CPU, unlike the In-Order CPU in the K1. This difference is crucial for performance. With in-order execution, the pipeline stalls on memory accesses; OoO, by contrast, allows independent instructions to proceed, boosting throughput and hardware utilization.

This effect is challenging to model and quantify in a general setting; instead, one must experiment with the specific algorithm on specific data sets and analyze various indirect performance indicators. We collected available PMU counters on the SpacemiT K1 and K3 using the \var{perf} utility. Although the set of available counters is relatively limited, we selected the most informative ones. For the experiments, we used matrix \var{kron\_g500-logn19} (524K rows, 524K columns, 44M nonzeros). \cref{tab:perf} shows the metrics for SpMV execution in three formats: CSR, Sell-C-$\sigma$, and HCSR.

\begin{table}[htbp] 
\centering
\caption{SpMV performance metrics on SpacemiT K1 and K3 for matrix kron\_g500-logn19.}
\label{tab:perf}

\footnotesize
\setlength{\tabcolsep}{4pt}
\begin{tabular}{lrrrr}
\toprule
\textbf{Metric} & \textbf{CSR Scalar} & \textbf{CSR RVV} & \textbf{Sell-C-$\sigma$} & \textbf{HCSR} \\
\midrule
\multicolumn{5}{c}{\textbf{SpacemiT K1}} \\
\midrule

Cycles (M)        & 51381  & 51739  & 70703  & 16591 \\
Instructions (M)  & 2190   & 527    & 2374   & 4017  \\
L1D misses (M)    & 219	   & 222    & 219    & 90    \\
Stalled cyc. (M)  & 49339  & 50561  & 67804  & 13513 \\
IPC               & 0.04   & 0.01   & 0.03   & 0.24  \\
Time (ms)         & 806    & 828    & 1110   & 258   \\
Eff. BW (GB/s)    & 1.08   & 1.05   & 0.78   & 3.38  \\
\midrule
\multicolumn{5}{c}{\textbf{SpacemiT K3}} \\
\midrule
Cycles (M)        & 8944   & 11280  & 30428 & 4377  \\
Instructions (M)  & 2093   & 429    & 2274  & 4006  \\
L1D misses (M)    & 207    & 205    & 207   & 87    \\
Stalled cyc. (M)  & 8412   & 11202  & 29759 & 3649  \\
IPC               & 0.23   & 0.04   & 0.07  & 0.92  \\
Time (ms)         & 102    & 131    & 348   & 49    \\
Eff. BW (GB/s)    & 8.53   & 6.64   & 2.50  & 17.63 \\
\bottomrule
\end{tabular}
\end{table}

The results show a significant performance gain on the K3 compared to the K1, consistent with the memory bandwidth improvement seen in the STREAM benchmark. The IPC values and stalled-cycles counter, which reflects the aggregate core idle time, confirm that the workload is memory-bound, with compute units stalled about 97\% and 83\% of time for standard and HCSR data structures, respectively, regardless of hardware.
The L1D-load-miss count, which measures L1 data cache misses, is almost the same for CSR and \textbf{Sell-C-$\sigma$} formats and does not change significantly when moving from K1 to K3. In contrast, the HCSR format exhibits a different behavior: since this data structure is designed to be cache-friendly in terms of vector element accesses, it performs much better due to the more efficient memory usage pattern.
 
Notably, OoO execution significantly affects SpMV performance and likely other sparse linear algebra algorithms. This stems from their low arithmetic intensity and irregular memory accesses, including indexed loads that lead to a large number of cache misses. The X100 OoO core in the SpacemiT K3 can partially hide these latencies by issuing and handling significantly more memory requests concurrently. This architectural feature explains the significant performance gain on the new hardware.

When comparing various sparse matrix storage formats, it is informative to examine their effective memory bandwidth. For our estimation, we assume that each nonzero element (NZ) requires loading at least 20 bytes of data in sparse formats. The effective memory bandwidth was calculated as the ratio of the useful data volume (20 bytes $\times$ NZ) to the execution time of a single multiplication. The analysis shows that the standard CSR algorithm on both hardware platforms utilizes almost 15\% and 50\% of the peak DRAM bandwidth on K1 and K3, respectively (the remainder corresponds to the inherent overhead of the algorithm). The \textbf{Sell-C-$\sigma$} format and vectorized CSR achieve lower effectively memory bandwidth for the matrix under study. This indicates that vectorized versions likely suffer from redundant data fetches. \textbf{HCSR} demonstrate near-peak DRAM bandwidth on the SpacemiT K3. Our estimates prove that its performance for this matrix is limited only by loading the nonzeros and index array. This is a strong indication of the algorithm's efficiency for SpMV.
Overall, the architectural changes introduced in the SpacemiT K3 have a substantial impact on SpMV performance across various sparse matrix formats, enabling a severalfold reduction in execution time for a broad range of sparse matrices, and likely for virtually all of them.

\section{Machine learning-based automatic format selection}
\label{sectionML}

Our experimental results confirm that SpMV execution time varies significantly depending on the chosen storage format and the characteristics of the matrix. The efficiency of a particular format is primarily determined by the nonzero structure, which dictates the memory access patterns and, consequently, the efficiency of processor cache utilization. Selecting the optimal storage format is therefore a non-trivial task, as the decision depends on subtle features of the sparsity pattern as well as the underlying hardware characteristics. Consequently, automatic format selection for sparse matrices is of great practical importance.

This section addresses the following key questions: Which specific matrix parameters influence the selection of the optimal storage format? Do these parameters vary across different formats? And is it possible to identify a set of matrix profile characteristics that enables a clear format choice without relying on a complex machine learning model?

\subsection{Model and feature selection}

In this work, we implemented several machine learning algorithms to automatically select the optimal matrix format for minimizing SpMV execution time, building upon ideas from the WISE framework \cite{28}. We evaluated Random Forest and Gradient Boosting Decision Trees for both classification and regression tasks. Classification models directly predicted the best format. Regression models estimated the ratio of SpMV execution time in a given format relative to the CSR baseline, ultimately selecting the format with the lowest predicted ratio.

To construct the feature vector, each matrix was partitioned into approximately equal-sized blocks (2048 segments along both rows and columns). The extracted features included basic matrix dimensions (number of rows, columns, and nonzeros) and the distribution characteristics of nonzeros across rows (R), columns (C), tiles (T), row blocks (RB), and column blocks (CB). For each distribution, we computed the sample mean, variance, standard deviation, minimum and maximum values, nonzero count, Gini index, and $p$-ratio. The $p$-ratio is defined as the minimum fraction $p \in [0, 1]$ such that the largest sample elements, constituting a fraction $p$ of the total element count, account for at least a fraction $1-p$ of the total sum.

The following features were also considered:
\begin{itemize}[itemsep=2pt, parsep=0pt, topsep=2pt, partopsep=0pt]
\item \textsf{uniqR} (\textsf{uniqC}): the average number of distinct row (column) indices per block;
\item \textsf{GrX\_uniqR} (\textsf{GrX\_uniqC}): analogous to \textsf{uniqR} (\textsf{uniqC}), but operating on groups of $X$ consecutive rows (columns);
\item \textsf{potReuseR} (\textsf{potReuseC}): the average number of blocks containing elements from a specific row (column);
\item \textsf{GrX\_potReuseR} (\textsf{GrX\_potReuseC}): analogous to \linebreak \textsf{potReuseR} (\textsf{potReuseC}), but operating on groups of $X$ rows (columns), with averages computed over the total number of corresponding groups.
\end{itemize}

We evaluated $X \in \{4, 8, 16, 32, 64\}$. These distribution characteristics capture computation balancing during row-parallelization, the specifics of accessing the $b$ vector, and spatial locality across different blocks. Specifically, the latter group of features reflects data access locality across various cache levels.

All implemented formats were included in the training phase; for parameterized formats, several fixed parameter combinations were evaluated. The dataset comprised 121 SuiteSparse matrices (used in our earlier experiments) and 600 R-MAT matrices \cite{48}. Models were trained using 8-thread execution data from both SpacemiT K1 and K3 boards. We used a random 80\% split for training, while testing was restricted exclusively to the SuiteSparse matrices. The efficiency of the automatic format selection was evaluated based on the speedup achieved over the CSR baseline.

\subsection{Experimental Results}
\begin{figure*}[t]
\centering
\includegraphics[width=\textwidth]{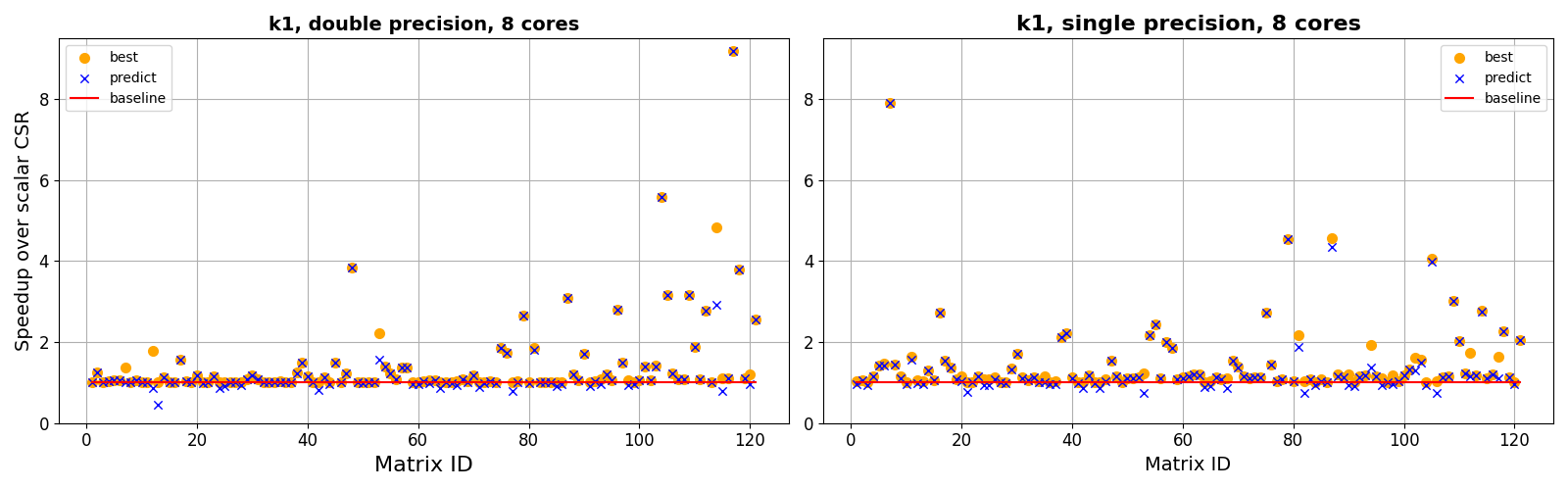}
\caption{SpMV operation speedup relative to scalar CSR when using the automatically chosen format (crosses) and the best of experiments (dots) on SpacemiT K1.} \label{k1-predict}
\end{figure*}

\begin{figure*}[t]
\centering
\includegraphics[width=\textwidth]{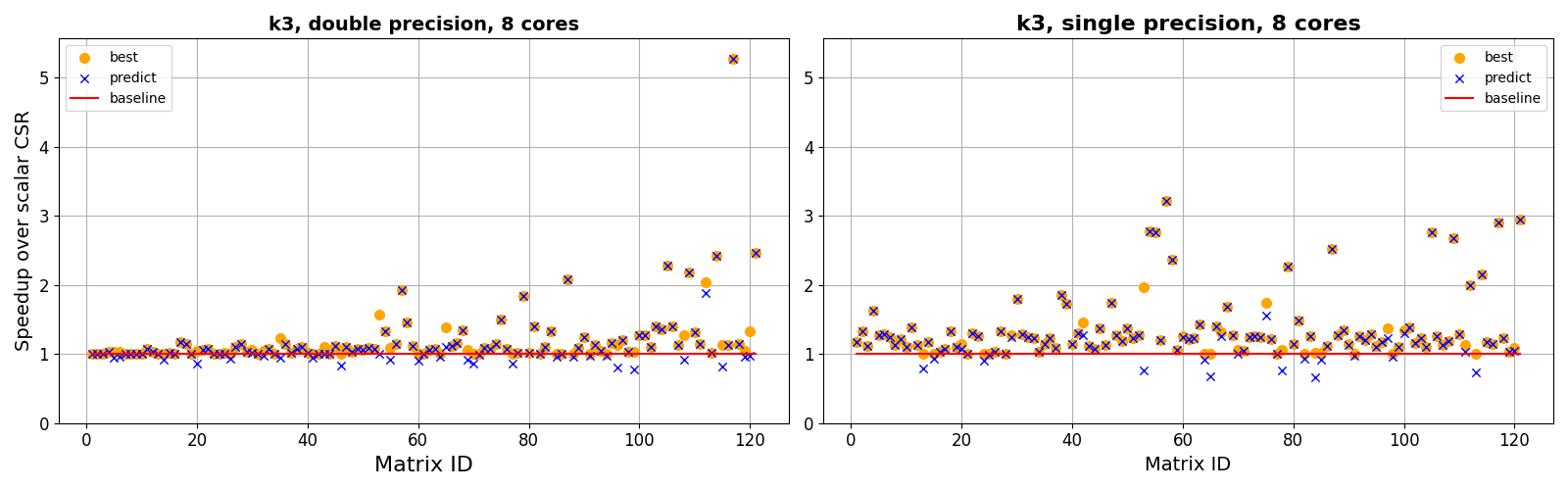}
\caption{SpMV operation speedup relative to scalar CSR when using the automatically chosen format (crosses) and the best of experiments (dots) on SpacemiT K3.} \label{k3-predict}
\end{figure*}

The results are presented in \cref{k1-predict} and \cref{k3-predict}. Relative speedup is shown only for the machine learning methods that demonstrated the best results. For both K1 and K3 platforms, this is RandomForestRegressor.

Using the model-predicted format for SpMV instead of scalar CSR yields an average speedup of of $1.26\times$ for single precision and $1.12\times$ for double precision on the SpacemiT K1 processor, and $1.25\times$ for single precision and $1.2\times$ for double precision on the SpacemiT K3 processor. In many cases, speedups of $2.0\times$--$5.0\times$ are achieved. In a small number of cases the selected format turns out to be slower than CSR.

The accuracy of format selection can be estimated by the loss under best (LUB), which is the ratio of the speedup from the optimal format to the speedup from the selected format. To improve the accuracy of format selection, hyperparameters for the machine learning methods were tuned. For GradientBoosting, these were num\_estimators, max\_depth, min\_samples\_leaf, and min\_samples\_split. The hyperparameter selection criterion was minimizing the maximum LUB across all matrices. As a result, we achieved a maximum LUB within 1.2–1.4 in all cases. However, the resulting models proved to be sensitive to the data: minor adjustments to the experimental results led to a significant deterioration in the accuracy of format selection in the worst case. On K3, for single precision, the LUB is 1.03 on average and 3.2 in the worst case, for double precision – 1.03 on average and 2.2 in the worst case. On K1, for single precision, the LUB is 1.05 on average and 2.2 in the worst case, for double precision – 1.04 on average and 2.1 in the worst case. In all cases where a large difference is observed, the HCSR format is predicted, but the optimal format turns out to be different.

For the formats that demonstrated the best performance, we analyzed which matrix features had the greatest impact on SpMV efficiency compared to CSR format. For each format, a numeric vector of feature importance was obtained from a random forest regression model trained exclusively on the experimental results for that format. 
For the formats that demonstrated the best performance, we analyzed which matrix features had the greatest impact on SpMV efficiency relative to the baseline CSR format. For each format, a feature importance vector was extracted using a random forest regression model trained strictly on the experimental results for that specific format. Higher values indicate greater feature importance, with all scores normalized to sum to 1.0.

The most important features for each format are presented in Supplementary Tables S2 and S3, separately for SpacemiT K1 and SpacemiT K3.

Based on the feature importance analysis, the following conclusions can be drawn:
\begin{enumerate}[itemsep=2pt, parsep=0pt, topsep=2pt, partopsep=0pt]
\item Significant differences in results are observed between K1 and K3, as well as between single and double precision. This reflects differences in the machine microarchitecture when using data types of different lengths. The results on K3 may be more realistic for future machines that will also implement out-of-order instruction execution.
\item It is typical that many formats have three or more features that are considered the most important, but each of them has a negligible importance, less than 0.1. This means that the features are quite complexly interrelated, making it very difficult to draw clear conclusions in such cases.
\item For the CSR5, CVR, and VNEC formats, some of the most features overlap, and they're related to irregular matrices.
\item For the SELL-C-$\sigma$ format, the most important features are not only from row distribution, which describes regulatiry of matrices, but also from column and tile distribution.
\item VHCC performance depends significantly on cache locality features and distribution of nonzeros over columns and column blocks.
\item HCSR performance is determined by cache locality features.
\end{enumerate}

An analysis of the influence of matrix features suggests that for many formats, features corresponding to the matrix types for which they were originally designed were significant. This confirms the adequacy of the model. However, it is difficult to trace the relationships between features in more detail. It is not yet possible to define a set of simple rules for selecting the optimal format when calling the SpMV function; a more complex procedure is required. The currently developed functionality for automatic selection of the optimal format can be used as an auxiliary tool, offering a fairly good, but not the best, solution.

\section{Conclusion}
\label{secConcl}
Despite numerous efforts by the research community, many algorithms operating on sparse matrices still fail to achieve even tens of percent of the peak performance of modern computing systems. SpMV is a prominent representative of memory‑bound algorithms, combining several factors that prevent efficient utilization of processor capabilities. First, its arithmetic intensity is low, since we perform only two floating‑point operations while loading two floating‑point numbers and an integer index from memory. Second, the irregular sparsity pattern leads to numerous cache misses and makes memory access remarkably inefficient. Third, the irregular distribution of nonzero entries may cause load imbalance in parallel execution. Many of these factors are inherent to the nature of sparse matrices and can hardly be eliminated entirely; nevertheless, developing new data structures and efficient implementations of the algorithm for novel architectures, which can accelerate computations at least to some extent, has clear practical value. This work is a step in this direction.

We have proposed a new format, HCSR, which is not only sufficiently simple in its design, inheriting numerous advantages of standard formats, but also improves memory access efficiency for a wide range of sparse matrices. Experiments demonstrated that the results of the baseline SpMV implementation in CSR on RISC‑V are consistent with those reported by other researchers and can be considered a fair reference. Based on this, we observed an average speedup of 19\% (13\% for K3) for double precision and 36\% (34\% for K3) for single precision when using HCSR on 8 cores of both K1 and K3 compared to the best CSR implementation. 

Furthermore, we have implemented support for nine popular sparse formats for the SpMV operation in our publicly available open‑source library RVVLASparse and performed a detailed performance comparison of SpMV in different formats on two generations of RISC‑V devices – SpacemiT K1 and SpacemiT K3. Particularly noteworthy are the changes we observed on the new K3 boards. The presence of an out‑of‑order processor enabled a severalfold acceleration of computations, which is not observed when working with dense matrices.

Experiments show that there is a group of sparse formats that dominate others on the vast majority of matrices. At the same time, selecting the most suitable format requires significant engineering effort. We adapted the methodology developed by the authors of WISE to our implementation and trained machine learning models that can select the most appropriate format for a given matrix structure. Additionally, we identified which structural properties of matrices influence the suitability of each format.

We hope that our results will be useful to other researchers when adapting algorithms to the RISC‑V architecture. In future work, we plan to focus on extending the range of supported sparse operations, as well as integrating our developments into iterative solvers for sparse linear systems.

\section*{Declaration of Generative AI and AI-assisted technologies in the writing process}
During the preparation of this work, the authors used AI tools in order to translate the paper, improve the English language phrasing and proofread the manuscript. After using AI tools, the authors reviewed and edited the manuscript and take full responsibility for the content of the publication.

\section*{CRediT authorship contribution statement}
Anna Pirova: Writing – original draft, Writing – review \& editing, Visualization, Validation, Software, Investigation, Formal analysis, Conceptualization, Project administration. 
Anastasia Vodeneeva: Writing – original draft, Visualization, Validation, Software, Investigation. 
Konstantin Kovalev: Visualization, Validation, Software. 
Alexander Ustinov: Writing – original draft, Visualization, Validation, Software. 
Maksim Zagriadskov: Visualization, Validation, Software.
Daniil Litvyakov: Visualization, Validation, Software.
Arthur Kulik: Visualization, Validation, Software.
Evgeny Kozinov: Visualization, Validation, Software, Investigation. 
Valentin Volokitin: Writing – review \& editing, Writing – original draft, Validation, Software, Methodology, Investigation, Formal analysis, Conceptualization. 
Iosif Meyerov: Writing – original draft, Writing – review \& editing, Visualization, Validation, Supervision, Project administration, Methodology, Investigation, Funding acquisition, Formal analysis, Conceptualization.

\section*{Declaration of competing interest}
The authors declare that they have no known competing financial interests or personal relationships that could have appeared to influence the work reported in this paper.

\section*{Acknowledgements}
The project is supported by the Lobachevsky University academic excellence program \enquote{Priority-2030}. The authors acknowledge the use of computational resources provided by the University (Lobachevsky Supercomputer).

\section*{Data availability} 
RVVLASparse is open-source software. The library is publicly available at \href{URL}{https://github.com/UNN-ITMM-Software/\-RVV\-LASparse}.

\begingroup
\setlength{\itemsep}{0pt} 
\bibliographystyle{elsarticle-num} 
\bibliography{refs.bib}
\endgroup

\onecolumn
\section{Supplementary material}
\renewcommand{\thetable}{S\arabic{table}}
\setcounter{table}{0}

\subsection{Test dataset}
\label{app1}

The test set consisted of 121 matrices from the SuiteSparse Matrix collection (\url{https://sparse.tamu.edu/}). Detailed information is given below (Table S1).

\begin{longtable}[!t]{r l r r r r r}
\caption{Matrix characteristics}
\label{tab:matrices} \\
\toprule
ID & Matrix name & \multicolumn{1}{c}{$m$} & \multicolumn{1}{c}{$n$} & \multicolumn{1}{c}{$nz$} & \multicolumn{1}{c}{$nz/row$} & \multicolumn{1}{c}{fill-in} \\
\midrule
\endfirsthead

\multicolumn{7}{c}{{\tablename\ \thetable{} -- continued from previous page}} \\
\toprule
ID & Matrix name & \multicolumn{1}{c}{$m$} & \multicolumn{1}{c}{$n$} & \multicolumn{1}{c}{$nz$} & \multicolumn{1}{c}{$nz/row$} & \multicolumn{1}{c}{fill-in} \\
\midrule

\endhead
\bottomrule
\multicolumn{7}{r}{{Continued on next page}} \\
\endfoot

\bottomrule
\endlastfoot

1   & italy\_osm          & 6,686,493 & 6,686,493 & 14,027,956  & 2 & 3.14e-07 \\
2   & Hardesty3           & 8,217,820 & 7,591,564 & 40,451,632  & 5 & 6.48e-07 \\
3   & delaunay\_n23       & 8,388,608 & 8,388,608 & 50,331,568  & 6 & 7.15e-07 \\
4   & rajat31             & 4,690,002 & 4,690,002 & 20,316,253  & 4 & 9.24e-07 \\
5   & patents             & 3,774,768 & 3,774,768 & 14,970,767  & 4 & 1.05e-06 \\
6   & cit-Patents         & 3,774,768 & 3,774,768 & 16,518,948  & 4 & 1.16e-06 \\
7   & circuit5M\_dc       & 3,523,317 & 3,523,317 & 14,865,409  & 4 & 1.20e-06 \\
8   & Freescale1          & 3,428,755 & 3,428,755 & 17,052,626  & 5 & 1.45e-06 \\
9   & Freescale2          & 2,999,349 & 2,999,349 & 14,313,235  & 5 & 1.59e-06 \\
10  & memchip             & 2,707,524 & 2,707,524 & 13,343,948  & 5 & 1.82e-06 \\
11  & cont11\_l           & 1,468,599 & 1,961,394 & 5,382,999   & 4 & 1.87e-06 \\
12  & cont1\_l            & 1,918,399 & 1,921,596 & 7,031,999   & 4 & 1.91e-06 \\
13  & circuit5M           & 5,558,326 & 5,558,326 & 59,524,291  & 11 & 1.93e-06 \\
14  & ljournal-2008       & 5,363,260 & 5,363,260 & 79,023,142  & 15 & 2.75e-06 \\
15  & delaunay\_n21       & 2,097,152 & 2,097,152 & 12,582,816  & 6 & 2.86e-06 \\
16  & LargeRegFile        & 2,111,154 & 801,374   & 4,944,201   & 2 & 2.92e-06 \\
17  & kkt\_power          & 2,063,494 & 2,063,494 & 12,771,361  & 6 & 3.00e-06 \\
18  & G3\_circuit         & 1,585,478 & 1,585,478 & 7,660,826   & 5 & 3.05e-06 \\
19  & webbase-1M          & 1,000,005 & 1,000,005 & 3,105,536   & 3 & 3.11e-06 \\
20  & rgg\_n\_2\_22\_s0   & 4,194,304 & 4,194,304 & 60,718,396  & 14 & 3.45e-06 \\
21  & indochina-2004      & 7,414,866 & 7,414,866 & 194,109,311 & 26 & 3.53e-06 \\
22  & channel-500x100x100-b050 & 4,802,000 & 4,802,000 & 85,362,744  & 18 & 3.70e-06 \\
23  & cage15              & 5,154,859 & 5,154,859 & 99,199,551  & 19 & 3.73e-06 \\
24  & Delor295K           & 295,734   & 1,823,928 & 2,401,323   & 8 & 4.45e-06 \\
25  & stormG2\_1000       & 528,185   & 1,377,306 & 3,459,881   & 7 & 4.76e-06 \\
26  & ecology1            & 1,000,000 & 1,000,000 & 4,996,000   & 5 & 5.00e-06 \\
27  & atmosmodd           & 1,270,432 & 1,270,432 & 8,814,880   & 7 & 5.46e-06 \\
28  & ASIC\_680k          & 682,862   & 682,862   & 2,638,997   & 4 & 5.66e-06 \\
29  & NACA0015            & 1,039,183 & 1,039,183 & 6,229,636   & 6 & 5.77e-06 \\
30  & web-Google          & 916,428   & 916,428   & 5,105,039   & 6 & 6.08e-06 \\
31  & vas\_stokes\_4M     & 4,382,246 & 4,382,246 & 131,577,616 & 30 & 6.85e-06 \\
32  & packing-500x100x100-b050 & 2,145,852 & 2,145,852 & 34,976,486  & 16 & 7.59e-06 \\
33  & mc2depi             & 525,825   & 525,825   & 2,100,225   & 4 & 7.60e-06 \\
34  & watson\_2           & 352,013   & 677,224   & 1,846,391   & 5 & 7.74e-06 \\
35  & in-2004             & 1,382,908 & 1,382,908 & 16,917,053  & 12 & 8.84e-06 \\
36  & Transport           & 1,602,111 & 1,602,111 & 23,487,281  & 15 & 9.15e-06 \\
37  & StocF-1465          & 1,465,137 & 1,465,137 & 21,005,389  & 14 & 9.78e-06 \\
38  & IMDB                & 428,440   & 896,308   & 3,782,463   & 9 & 9.85e-06 \\
39  & soc-Pokec           & 1,632,803 & 1,632,803 & 30,622,564  & 19 & 1.15e-05 \\
40  & cage14              & 1,505,785 & 1,505,785 & 27,130,349  & 18 & 1.20e-05 \\
41  & ss                  & 1,652,680 & 1,652,680 & 34,753,577  & 21 & 1.27e-05 \\
42  & Delor338K           & 343,236   & 887,058   & 4,211,599   & 12 & 1.38e-05 \\
43  & vas\_stokes\_2M     & 2,146,677 & 2,146,677 & 65,129,037  & 30 & 1.41e-05 \\
44  & Hardesty2           & 929,901   & 303,645   & 4,020,731   & 4 & 1.42e-05 \\
45  & Bump\_2911          & 2,911,419 & 2,911,419 & 127,729,899 & 44 & 1.51e-05 \\
46  & Stanford\_Berkeley  & 683,446   & 683,446   & 7,583,376   & 11 & 1.62e-05 \\
47  & amazon0601          & 403,394   & 403,394   & 3,387,388   & 8 & 2.08e-05 \\
48  & af\_shell10         & 1,508,065 & 1,508,065 & 52,259,885  & 35 & 2.30e-05 \\
49  & nlpkkt80            & 1,062,400 & 1,062,400 & 28,192,672  & 27 & 2.50e-05 \\
50  & Hook\_1498          & 1,498,023 & 1,498,023 & 59,374,451  & 40 & 2.65e-05 \\
51  & Cube\_Coup\_dt0     & 2,164,760 & 2,164,760 & 124,406,070 & 57 & 2.66e-05 \\
52  & Cube\_Coup\_dt6     & 2,164,760 & 2,164,760 & 124,406,070 & 57 & 2.66e-05 \\
53  & ins2                & 309,412   & 309,412   & 2,751,484   & 9 & 2.87e-05 \\
54  & Stanford            & 281,903   & 281,903   & 2,312,497   & 8 & 2.91e-05 \\
55  & web-Stanford        & 281,903   & 281,903   & 2,312,497   & 8 & 2.91e-05 \\
56  & vas\_stokes\_1M     & 1,090,664 & 1,090,664 & 34,767,207  & 32 & 2.92e-05 \\
57  & NotreDame\_actors   & 392,400   & 127,823   & 1,470,404   & 4 & 2.93e-05 \\
58  & mac\_econ\_fwd500   & 206,500   & 206,500   & 1,273,389   & 6 & 2.99e-05 \\
59  & Serena              & 1,391,349 & 1,391,349 & 64,131,971  & 46 & 3.31e-05 \\
60  & Rucci1              & 1,977,885 & 109,900   & 7,791,168   & 4 & 3.58e-05 \\
61  & dielFilterV2real    & 1,157,456 & 1,157,456 & 48,538,952  & 42 & 3.62e-05 \\
62  & Long\_Coup\_dt6     & 1,470,152 & 1,470,152 & 84,422,970  & 57 & 3.91e-05 \\
63  & Flan\_1565          & 1,564,794 & 1,564,794 & 114,165,372 & 73 & 4.66e-05 \\
64  & ldoor               & 952,203   & 952,203   & 42,493,817  & 45 & 4.69e-05 \\
65  & turon\_m            & 189,924   & 189,924   & 1,690,876   & 9 & 4.69e-05 \\
66  & Emilia\_923         & 923,136   & 923,136   & 40,373,538  & 44 & 4.74e-05 \\
67  & ML\_Geer            & 1,504,002 & 1,504,002 & 110,686,677 & 74 & 4.89e-05 \\
68  & bone010             & 986,703   & 986,703   & 47,851,783  & 48 & 4.92e-05 \\
69  & sls                 & 1,748,122 & 62,729    & 6,804,304   & 4 & 6.21e-05 \\
70  & gsm\_106857         & 589,446   & 589,446   & 21,758,924  & 37 & 6.27e-05 \\
71  & degme               & 185,501   & 659,415   & 8,127,528   & 44 & 6.64e-05 \\
72  & Fault\_639          & 638,802   & 638,802   & 27,245,944  & 43 & 6.67e-05 \\
73  & PFlow\_742          & 742,793   & 742,793   & 37,138,461  & 50 & 6.73e-05 \\
74  & af\_2\_k101         & 503,625   & 503,625   & 17,550,675  & 35 & 6.92e-05 \\
75  & higgs-twitter       & 456,626   & 456,626   & 14,855,842  & 33 & 7.13e-05 \\
76  & dielFilterV3real    & 1,102,824 & 1,102,824 & 89,306,020  & 81 & 7.34e-05 \\
77  & bundle\_adj         & 513,351   & 513,351   & 20,207,907  & 39 & 7.66e-05 \\
78  & tp-6                & 142,752   & 1,014,301 & 11,537,419  & 81 & 7.97e-05 \\
79  & kron\_g500-logn20   & 1,048,576 & 1,048,576 & 89,239,674  & 85 & 8.11e-05 \\
80  & audikw\_1           & 943,695   & 943,695   & 77,651,847  & 82 & 8.72e-05 \\
81  & hollywood-2009      & 1,139,905 & 1,139,905 & 113,891,327 & 100 & 8.76e-05 \\
82  & stat96v3            & 33,841    & 1,113,780 & 3,317,736   & 98 & 8.81e-05 \\
83  & CoupCons3D          & 416,800   & 416,800   & 17,277,420  & 41 & 9.95e-05 \\
84  & stat96v2            & 29,089    & 957,432   & 2,852,184   & 98 & 1.02e-04 \\
85  & msdoor              & 415,863   & 415,863   & 19,173,163  & 46 & 1.11e-04 \\
86  & inline\_1           & 503,712   & 503,712   & 36,816,170  & 73 & 1.45e-04 \\
87  & kron\_g500-logn19   & 524,288   & 524,288   & 43,562,265  & 83 & 1.58e-04 \\
88  & Si87H76             & 240,369   & 240,369   & 10,661,631  & 44 & 1.84e-04 \\
89  & ML\_Laplace         & 377,002   & 377,002   & 27,582,698  & 73 & 1.94e-04 \\
90  & BenElechi1          & 245,874   & 245,874   & 13,150,496  & 53 & 2.18e-04 \\
91  & F1                  & 343,791   & 343,791   & 26,837,113  & 78 & 2.27e-04 \\
92  & pwtk                & 217,918   & 217,918   & 11,524,432  & 53 & 2.43e-04 \\
93  & halfb               & 224,617   & 224,617   & 12,387,821  & 55 & 2.45e-04 \\
94  & Ga41As41H72         & 268,096   & 268,096   & 18,488,476  & 69 & 2.57e-04 \\
95  & troll               & 213,453   & 213,453   & 11,985,111  & 56 & 2.63e-04 \\
96  & fcondp2             & 201,822   & 201,822   & 11,294,316  & 56 & 2.77e-04 \\
97  & karted              & 46,502    & 133,115   & 1,770,349   & 38 & 2.86e-04 \\
98  & fullb               & 199,187   & 199,187   & 11,708,077  & 59 & 2.95e-04 \\
99  & bmw7st\_1           & 141,347   & 141,347   & 7,318,399   & 52 & 3.66e-04 \\
100 & gearbox             & 153,746   & 153,746   & 9,080,404   & 59 & 3.84e-04 \\
101 & Si41Ge41H72         & 185,639   & 185,639   & 15,011,265  & 81 & 4.35e-04 \\
102 & s4dkt3m2            & 90,449    & 90,449    & 3,753,461   & 41 & 4.59e-04 \\
103 & SiO2                & 155,331   & 155,331   & 11,283,503  & 73 & 4.68e-04 \\
104 & ESOC                & 327,062   & 37,830    & 6,019,939   & 18 & 4.87e-04 \\
105 & kron\_g500-logn17   & 131,072   & 131,072   & 10,228,360  & 78 & 5.95e-04 \\
106 & torso1              & 116,158   & 116,158   & 8,516,500   & 73 & 6.31e-04 \\
107 & pkustk14            & 151,926   & 151,926   & 14,836,504  & 98 & 6.43e-04 \\
108 & x104                & 108,384   & 108,384   & 8,713,602   & 80 & 7.42e-04 \\
109 & 12month1            & 12,471    & 872,622   & 22,624,727  & 1814 & 2.08e-03 \\
110 & JP                  & 87,616    & 67,320    & 13,734,559  & 157 & 2.33e-03 \\
111 & mip1                & 66,463    & 66,463    & 10,352,819  & 156 & 2.34e-03 \\
112 & rail4284            & 4,284     & 1,096,894 & 11,284,032  & 2634 & 2.40e-03 \\
113 & raefsky3            & 21,200    & 21,200    & 1,488,768   & 70 & 3.31e-03 \\
114 & rail2586            & 2,586     & 923,269   & 8,011,362   & 3098 & 3.36e-03 \\
115 & crankseg\_2         & 63,838    & 63,838    & 14,148,858  & 222 & 3.47e-03 \\
116 & nd24k               & 72,000    & 72,000    & 28,715,634  & 399 & 5.54e-03 \\
117 & connectus           & 512       & 394,792   & 1,127,525   & 2202 & 5.58e-03 \\
118 & specular            & 477,976   & 1,600     & 7,647,040   & 16 & 1.00e-02 \\
119 & nd12k               & 36,000    & 36,000    & 14,220,946  & 395 & 1.10e-02 \\
120 & TSOPF\_RS\_b2383    & 38,120    & 38,120    & 16,171,169  & 424 & 1.11e-02 \\
121 & spal\_004           & 10,203    & 321,696   & 46,168,124  & 4525 & 1.41e-02 \\
\bottomrule
\end{longtable}

\newpage
\subsection{Matrix portrait features analysis}
\label{app2}

The key matrix portrait features for each sparse format are presented in Tables S2 and S3 for SpacemiT K1 and K3, respectively. 

\begin{table*}[!htb]
\centering
\caption{The most important matrix portrait features for each format. SpacemiT K1.}
\label{tab:k1_features}
\begin{tabularx}{\textwidth}{l X X}
\toprule
\textbf{Sparse format} & \multicolumn{1}{c}{\textbf{Double precision}} & \multicolumn{1}{c}{\textbf{Single precision}} \\
\midrule
\textbf{CSRvec} & 
    \begin{itemize}[nosep, leftmargin=*] 
        \item RB-distribution, $p$-ratio
        \item R-distribution, $p$-ratio
        \item C-distribution, $p$-ratio
        \item CB-distribution, min
        \item C-distribution, max
    \end{itemize} &
    \begin{itemize}[nosep, leftmargin=*]
        \item RB-distribution, $p$-ratio
        \item R-distribution, $p$-ratio
        \item C-distribution, $p$-ratio
        \item CB-distribution, min
        \item C-distribution, max
    \end{itemize} \\
\midrule
\textbf{CSR5} &
    \begin{itemize}[nosep, leftmargin=*]
        \item Gr64\_potReuseC
        \item Number of columns
        \item R-distribution, nonzeros
        \item T-distribution, Gini index
        \item R-distribution, variance
    \end{itemize} &
    \begin{itemize}[nosep, leftmargin=*]
        \item RB-distribution, Gini index
        \item Gr4\_potReuseR
        \item R-distribution, $p$-ratio
        \item Gr32\_uniqR
        \item CB-distribution, variance
    \end{itemize} \\
\midrule
\textbf{CVR} &
    \begin{itemize}[nosep, leftmargin=*]
        \item Number of rows
        \item R-distribution, p-ratio
        \item CB-distribution, min
        \item C-distribution, p-ratio
    \end{itemize} &
    \begin{itemize}[nosep, leftmargin=*]
        \item Gr4\_potReuseR
        \item Gr8\_potReuseR
        \item R-distribution, Gini index
        \item R-distribution, variance
        \item T-distribution, Gini index
        \item T-distribution, $p$-ratio
    \end{itemize} \\
\midrule
\textbf{Sell-C-$\sigma$} &
    \begin{itemize}[nosep, leftmargin=*]
        \item R-distribution, Gini index
        \item R-distribution, p-ratio
        \item C-distribution, max
        \item C-distribution, p-ratio
    \end{itemize} &
    \begin{itemize}[nosep, leftmargin=*]
        \item R-distribution, Gini index
        \item R-distribution, p-ratio
        \item C-distribution, max
        \item C-distribution, $p$-ratio
    \end{itemize} \\
\midrule
\textbf{VHCC} &
    \begin{itemize}[nosep, leftmargin=*]
        \item RB-distribution, p-ratio
        \item CB-distribution, min
        \item C-distribution, p-ratio
        \item R-distribution, Gini index
        \item C-distribution, max
    \end{itemize} &
    \begin{itemize}[nosep, leftmargin=*]
        \item uniqC
        \item Gr8\_uniqR
        \item Gr64\_uniqR
        \item Gr32\_uniqR
        \item T-distribution, $p$-ratio
    \end{itemize} \\
\midrule
\textbf{VNEC} &
    \begin{itemize}[nosep, leftmargin=*]
        \item RB-distribution, p-ratio
        \item CB-distribution, min
        \item R-distribution, Gini index
        \item C-distribution, p-ratio
        \item C-distribution, max
    \end{itemize} &
    \begin{itemize}[nosep, leftmargin=*]
        \item Gr8\_potReuseR
        \item Gr4\_potReuseR
        \item Gr4\_uniqC
        \item potReuseC
    \end{itemize} \\
\midrule
\textbf{HCSR} &
    \begin{itemize}[nosep, leftmargin=*]
        \item Gr4\_potReuseR
        \item Gr8\_potReuseR
        \item potReuseC
        \item Gr16\_potReuseC
    \end{itemize} &
    \begin{itemize}[nosep, leftmargin=*]
        \item Gr32\_potReuseR
        \item Gr8\_potReuseC
        \item Gr16\_potReuseC
        \item Gr64\_potReuseC
        \item Gr8\_uniqC
        \item Gr16\_potReuseR
    \end{itemize} \\
\bottomrule
\end{tabularx}
\end{table*}

\begin{table*}[!htb]
\centering
\caption{The most important matrix portrait features for each format. SpacemiT K3.}
\label{tab:k3_features}
\begin{tabularx}{\textwidth}{l X X}
\toprule
\textbf{Sparse format} & \multicolumn{1}{c}{\textbf{Double precision}} & \multicolumn{1}{c}{\textbf{Single precision}} \\
\midrule
\textbf{CSRvec} & 
    \begin{itemize}[nosep, leftmargin=*]
        \item uniqC
        \item C-distribution, max
        \item R-distribution, variance
        \item CB-distribution, min
        \item potReuseR
    \end{itemize} &
    \begin{itemize}[nosep, leftmargin=*]
        \item CB-distribution, min 
        \item R-distribution, p-ratio 
        \item C-distribution, p-ratio 
        \item C-distribution, max
    \end{itemize} \\
\midrule
\textbf{CSR5} & 
    \begin{itemize}[nosep, leftmargin=*]
        \item uniqC
        \item Gr32\_uniqR                    
        \item C-distribution, variance    
        \item Gr8\_uniqR                     
        \item R-distribution, variance
    \end{itemize} &
    \begin{itemize}[nosep, leftmargin=*]
        \item Gr32\_uniqR                    
        \item R-distribution, variance
        \item Gr64\_uniqR                    
        \item C-distribution, variance    
    \end{itemize} \\
\midrule
\textbf{CVR} &
    \begin{itemize}[nosep, leftmargin=*]
        \item Gr4\_potReuseR                 
        \item Gr16\_potReuseR                
        \item Gr8\_potReuseR                 
        \item Gr16\_potReuseC                
        \item R-distribution, variance
    \end{itemize} &
    \begin{itemize}[nosep, leftmargin=*]
        \item Gr32\_potReuseR                
        \item Gr4\_potReuseR                 
        \item Gr16\_potReuseR                
        \item Gr16\_uniqR                    
        \item R-distribution, variation
        \item C-distribution, mean        
    \end{itemize} \\
\midrule
\textbf{Sell-C-$\sigma$} &
    \begin{itemize}[nosep, leftmargin=*]
        \item R-distribution, Gini index  
        \item R-distribution, variance
        \item R-distribution, $p$-ratio     
        \item C-distribution, max     
        \item C-distribution, $p$-ratio     
        \item T-distribution, $p$-ratio    
    \end{itemize} &
    \begin{itemize}[nosep, leftmargin=*]
        \item R-distribution, $p$-ratio
        \item R-distribution, variance
        \item C-distribution, $p$-ratio
        \item C-distribution, max
        \item C-distribution, variance
    \end{itemize} \\
\midrule
\textbf{VHCC} &
    \begin{itemize}[nosep, leftmargin=*]
        \item C-distribution, max      
        \item Gr64\_potReuseC            
        \item C-disitrubtion, variance     
        \item CB-distribution, min
        \item CB-distribution, max
    \end{itemize} &
    \begin{itemize}[nosep, leftmargin=*]
        \item Gr16\_uniqC                
        \item T-distribution, nonzeros
        \item Gr64\_uniqR                        
        \item Number of nonzeros                               
    \end{itemize} \\
\midrule
\textbf{VNEC} &
    \begin{itemize}[nosep, leftmargin=*]
        \item C-distribution, max  
        \item Gr8\_potReuseC                    
        \item R-distribution, nonzeros
        \item Gr8\_uniqC                        
    \end{itemize} &
    \begin{itemize}[nosep, leftmargin=*]
        \item R-distribution, Gini index      
        \item C-distribution, max        
        \item Gr16\_uniqR               
        \item R-distribution, $p$-ratio         
        \item CB-distribution, Gini index
    \end{itemize} \\
\midrule
\textbf{HCSR} &
    \begin{itemize}[nosep, leftmargin=*]
        \item Gr8\_potReuseR 
        \item Gr4\_potReuseR 
        \item potReuseC     
        \item Gr16\_potReuseC
        \item Gr16\_uniqR    
    \end{itemize} &
    \begin{itemize}[nosep, leftmargin=*]
        \item Gr16\_uniqC                       
        \item T-distribution, $p$-ratio       
        \item Gr8\_uniqR                        
        \item Gr32\_potReuseR                   
        \item R-distribution, nonzeros
    \end{itemize} \\
\bottomrule
\end{tabularx}
\end{table*}

\end{document}